\documentclass[letterpaper]{article}

\usepackage{proceed2e}
\usepackage[margin=22.2mm]{geometry}
\usepackage[utf8]{inputenc}
\usepackage{amsmath}
\usepackage{amssymb}
\usepackage{amsthm}
\usepackage{mathrsfs}
\usepackage{accents}
\usepackage{booktabs}     
\usepackage{amsfonts}    
\usepackage{nicefrac}  
\usepackage{microtype}
\usepackage{stix}
\usepackage{upgreek}
\usepackage[subrefformat=parens]{subcaption}
\usepackage{times}
\usepackage{graphicx}
\usepackage[square,sort,comma,numbers]{natbib}
\usepackage{hyperref}
\usepackage{relsize}
\usepackage{tikz}
\usetikzlibrary{bayesnet,arrows,calc,positioning}
\usepackage{xcolor}
\definecolor{links}{RGB}{33, 97, 140}
\hypersetup{
    colorlinks,
    linkcolor=links,
    citecolor=links,
    urlcolor=links
}

\title{Continuously tempered Hamiltonian Monte Carlo}

\author{ 
{ \bf Matthew M.~Graham } \\
School of Informatics \\
University of Edinburgh \\
\And
{ \bf Amos J.~Storkey } \\
School of Informatics \\
University of Edinburgh \\
}

\DeclareMathOperator{\probdens}{\mathbb{p}}
\DeclareMathOperator{\probability}{\mathbb{P}}
\DeclareMathOperator{\expectation}{\mathbb{E}}

\DeclareMathOperator{\gvn}{|}
\DeclareMathOperator{\Tr}{Tr}

\DeclareMathOperator{\diag}{diag}

\let\originalleft\left
\let\originalright\right
\renewcommand{\left}{\mathopen{}\mathclose\bgroup\originalleft}
\renewcommand{\right}{\aftergroup\egroup\originalright}

\newcommand{\vct}[1]{\boldsymbol{#1}}
\newcommand{\mtx}[1]{\boldsymbol{#1}}
\newcommand{\set}[1]{\mathcal{#1}}
\newcommand{\fset}[1]{\lbr #1 \rbr}
\newcommand{\reals}{\mathbb{R}}

\newcommand{\lpa}{\left(}
\newcommand{\rpa}{\right)}
\newcommand{\lbr}{\left\lbrace}
\newcommand{\rbr}{\right\rbrace}
\newcommand{\lsb}{\left[}
\newcommand{\rsb}{\right]}
\newcommand{\prob}[1]{\probability\lsb #1 \rsb}
\newcommand{\pden}[1]{\probdens\lsb #1 \rsb}
\newcommand{\rvar}[1]{\mathsf{#1}}
\newcommand{\rvct}[1]{\boldsymbol{\rvar{#1}}}
\newcommand{\nrm}[1]{\mathcal{N}\lpa #1 \rpa}

\newcommand{\expc}[1]{\expectation\lsb #1 \rsb}

\newcommand{\dr}{\mathrm{d}}
\newcommand{\td}[2]{\frac{\dr #1}{\dr #2}}

\newcommand{\pd}[2]{\frac{\partial #1}{\partial #2}}

\newcommand{\tr}{^{\mkern-1.5mu\mathsf{T}}}

\newcommand{\hamiltonian}{h}
\newcommand{\extended}[1]{\skew{2}{\tilde}{#1}}
\newcommand{\normconst}{Z}
\newcommand{\partfunc}{z}
\newcommand{\transition}{\mathbb{T}}

\newcommand{\abbrdef}[2]{\textit{#1}~(\textrm{#2})}
\newcommand{\abbrref}[1]{\textrm{#1}}

\makeatletter 
  \g@addto@macro\@uclclist{%
    \eth\Eth
    \thorn\Thorn
    \alpha\Alpha
    \beta\Beta
    \gamma\Gamma
    \delta\Delta
    \epsilon\Epsilon
    \varepsilon\Varepsilon
    \zeta\Zeta
    \eta\Eta
    \theta\Theta
    \vartheta\Vartheta
    \iota\Iota
    \kappa\Kappa
    \lambda\Lambda
    \mu\Mu
    \nu\Nu
    \xi\Xi
    \omicron\Omicron
    \pi\Pi
    \varpi\Varpi
    \rho\Rho
    \varrho\Varrho
    \sigma\Sigma
    \varsigma\Varsigma
    \tau\Tau
    \upsilon\Upsilon
    \phi\Phi
    \varphi\Varphi
    \chi\Chi
    \psi\Psi
    \omega\Omega
  }

\begin{document}

\maketitle

\begin{abstract}
Hamiltonian Monte Carlo (HMC) is a powerful Markov chain Monte Carlo (MCMC) method for performing approximate inference in complex probabilistic models of continuous variables. In common with many MCMC methods, however, the standard HMC approach performs poorly in distributions with multiple isolated modes. We present a method for augmenting the Hamiltonian system with an extra continuous temperature control variable which allows the dynamic to bridge between sampling a complex target distribution and a simpler unimodal base distribution. This augmentation both helps improve mixing in multimodal targets and allows the normalisation constant of the target distribution to be estimated. The method is simple to implement within existing HMC code, requiring only a standard leapfrog integrator. We demonstrate experimentally that the method is competitive with annealed importance sampling and simulating tempering methods at sampling from challenging multimodal distributions and estimating their normalising constants.
\end{abstract}

\section{Introduction}

Applications of \abbrdef{Markov chain Monte Carlo}{MCMC} methods to perform inference in challenging statistical physics problems were among the earliest uses of the first modern electronic computers \citep{robert2011short}. In the years since, \abbrref{MCMC} methods have become a mainstay for performing approximate inference in complex probabilistic models in most fields of computational science. Statistical physics in particular has continued to play a key role in the development of \abbrref{MCMC} methodology, both directly through the wider adoption of advances originally developed within the field such as tempering methods \citep{swendsen1986replica,geyer1991markov,marinari1992simulated} and gradient-based dynamics \citep{duane1987hybrid}, and more generally as a source of physical intuition for methods such as Gibbs sampling \citep{geman1984stochastic,gelfand1990sampling}.

The aim in \abbrref{MCMC} methods is to construct a Markov chain which has as its unique stationary distribution a target distribution, such that samples from the chain can be used to form Monte Carlo estimates of expectations with respect to the target distribution. Several general purpose constructions have been developed for specifying valid Markov transition operators for arbitrary target distributions, including the seminal Metropolis--Hastings algorithm \citep{metropolis1953equation,hastings1970monte} as well alternatives such as slice \citep{neal2003slice,damlen1999gibbs} and Gibbs sampling \citep{geman1984stochastic,gelfand1990sampling}.

While these methods give a basis for constructing chains that will asymptotically give correct inferences, producing algorithms that will give reasonable results in a practical amount of time is still a major challenge. For the restricted case of target distributions defined by densities that are differentiable functions of a real-valued vector variable, \abbrdef{Hamiltonian Monte Carlo}{HMC} \citep{duane1987hybrid,neal2011mcmc}\footnote{Originally termed \emph{Hybrid Monte Carlo} in \citep{duane1987hybrid}.} provides a framework for defining Markov chains that use the gradient of the target density to converge quickly to the target distribution and to make efficient large moves around it once it is reached.

Although \abbrref{HMC} is able to efficiently explore contiguous regions of high probability density in a target distribution, as with most \abbrref{MCMC} methods using local moves it struggles to move between isolated modes in multimodal target distributions \citep{neal2011mcmc}. Tempering methods \citep{swendsen1986replica,geyer1991markov,marinari1992simulated} which augment the state space with an (inverse) temperature variable are often used to improve exploration in multimodal distributions. As the temperature variable is typically discrete however it cannot be updated with a gradient-based \abbrref{HMC} dynamic.

In this paper we present an alternative \emph{continuous tempering} approach which instead augments the state with a continuous temperature variable. This allows use of \abbrref{HMC} to jointly update both the temperature and original state, making the approach simple to use with existing \abbrref{HMC} implementations. The proposed approach both improves exploration of multimodal target densities and also provides estimates of the typically unknown normalising constant of the target density, which is often an important inferential quantity in its own right for model comparison purposes \citep{gelman1998simulating}.

\begin{figure*}[!ht]
\begin{subfigure}[b]{.5\linewidth}
\vskip 0pt
\centering
\includegraphics[width=0.95\textwidth]{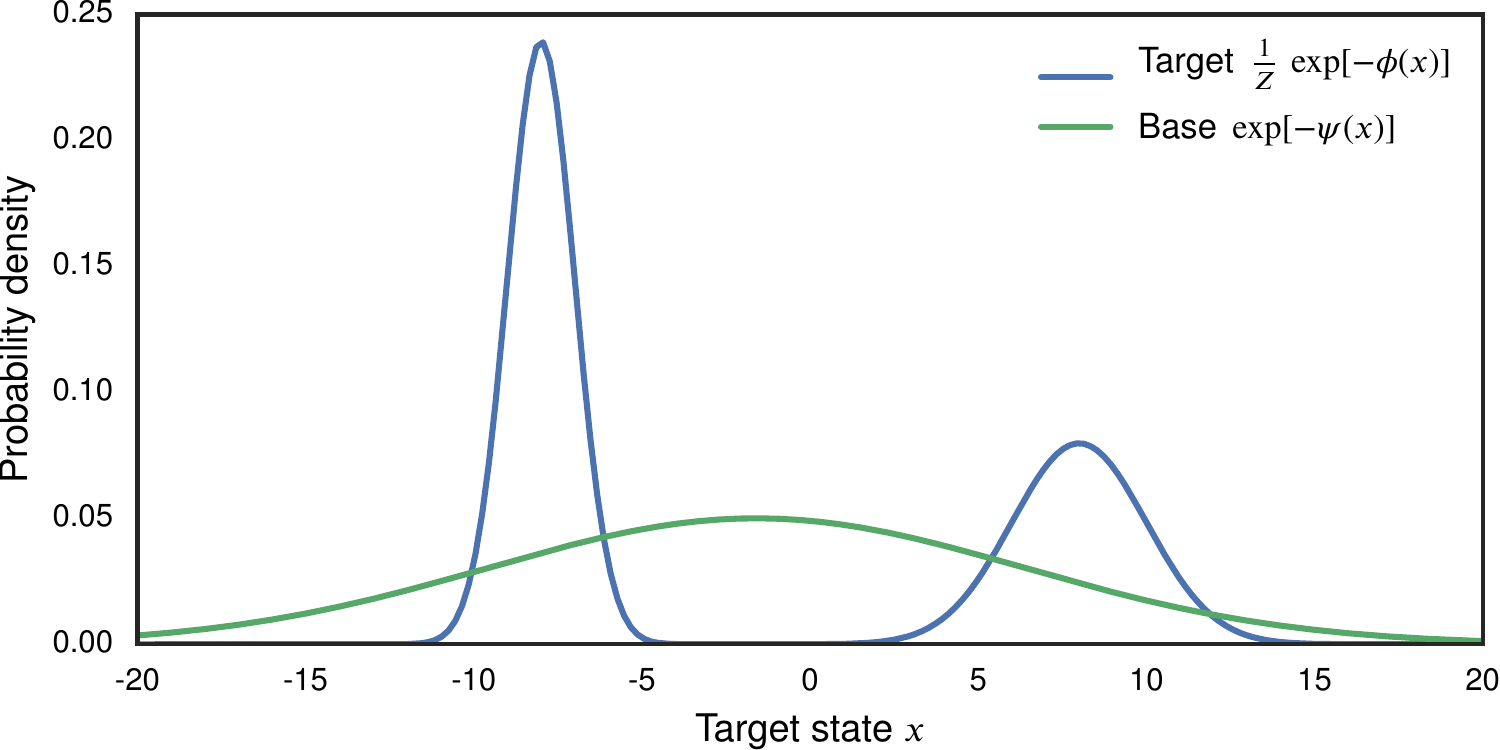} 
\caption{Target (blue curve) \& base (green curve) density functions.}\label{sfig:bimodal-gm-target-and-gaussian-base}
\end{subfigure}%
\begin{subfigure}[b]{.5\linewidth}
\vskip 0pt
\centering
\includegraphics[width=0.95\textwidth]{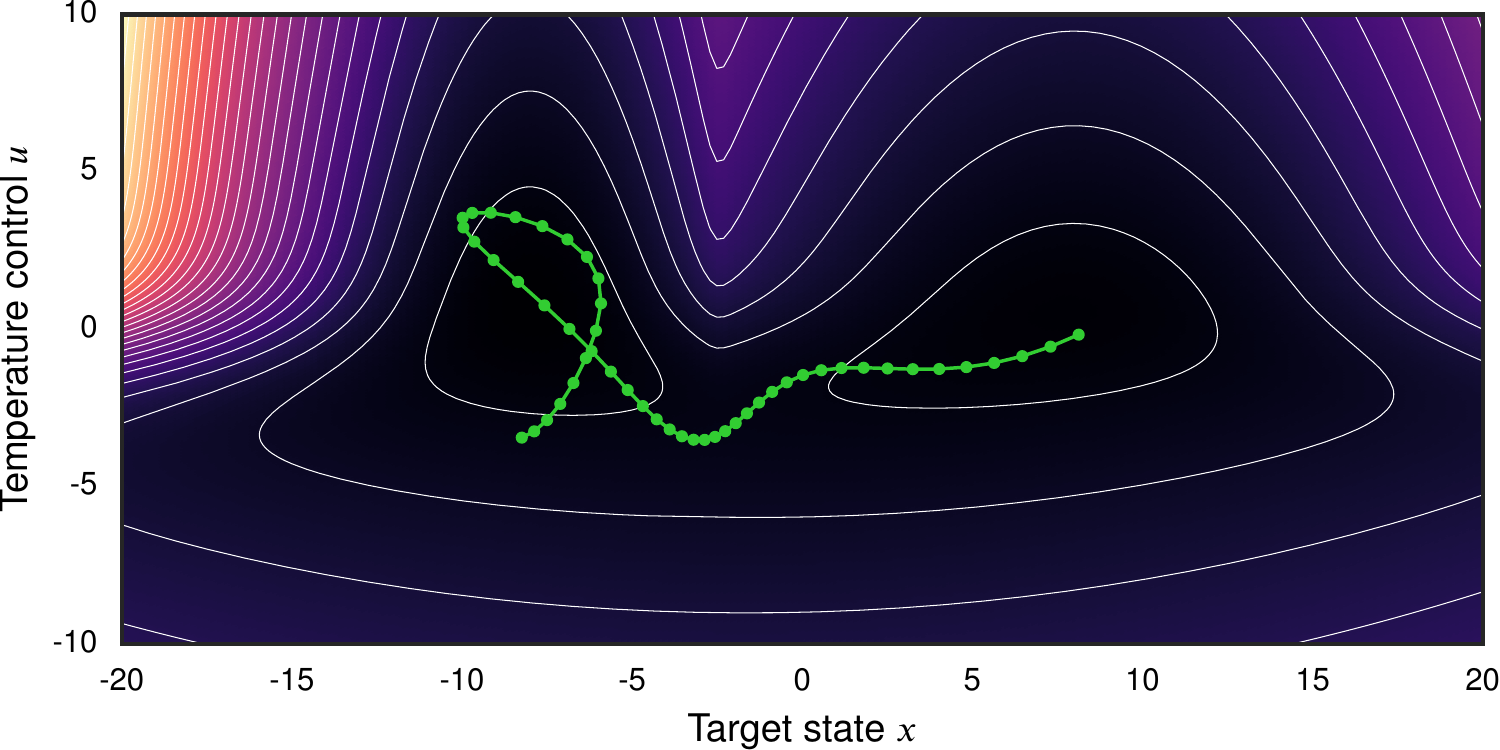} 
\caption{Joint energy (contour plot) \& example trajectory (green curve).}\label{sfig:jct-energy-and-trajectory}
\end{subfigure}%
\\
\begin{subfigure}[b]{.5\linewidth}
\vskip 5pt
\centering
\includegraphics[width=0.95\textwidth]{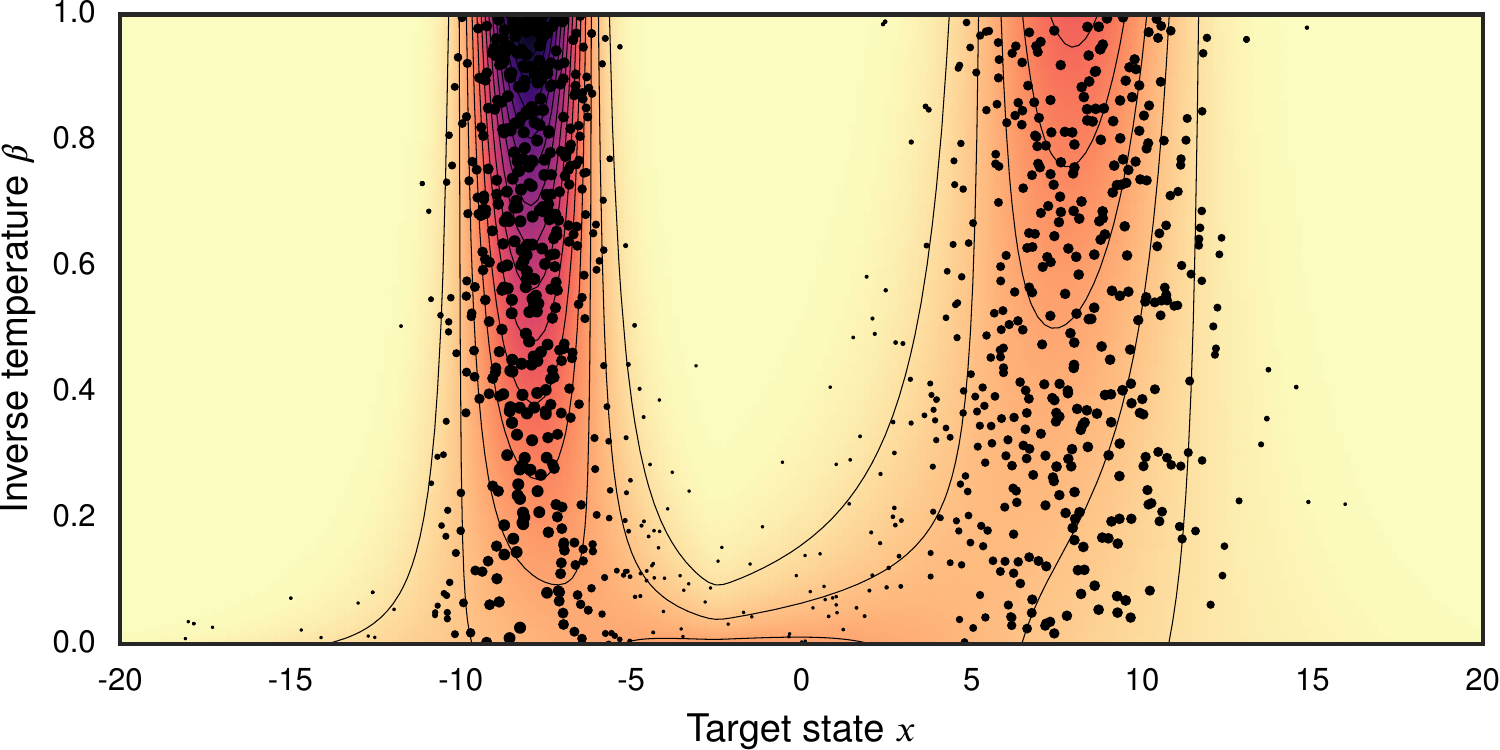}
\caption{Joint density (contour plot) and CT HMC samples (circles).}\label{sfig:jct-prob-dens-and-joint-samples}
\end{subfigure}
\begin{subfigure}[b]{.5\linewidth}
\vskip 5pt
\centering
\includegraphics[width=0.95\textwidth]{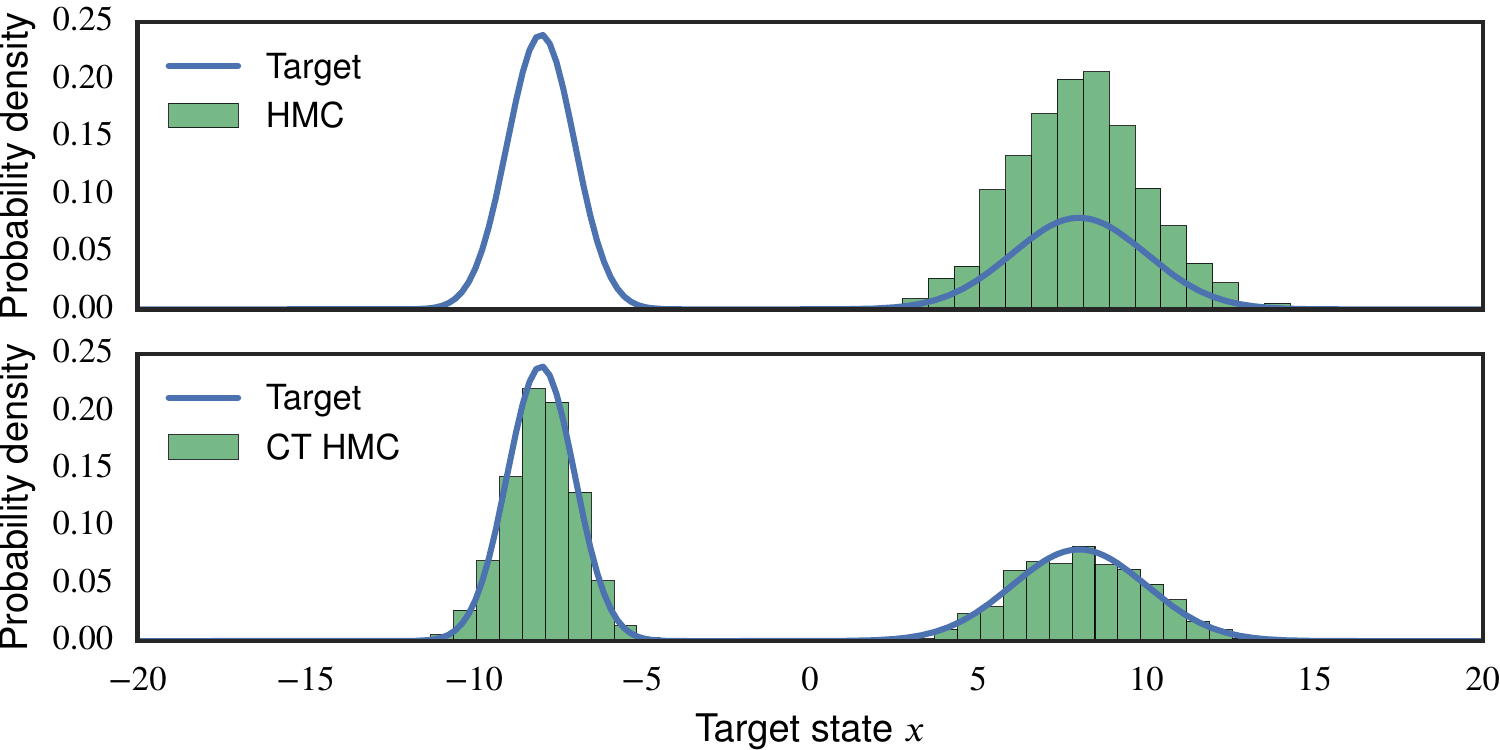}
\caption{Histograms from HMC (top) and CT HMC (bottom) samples.}\label{sfig:jct-and-hmc-target-histograms}
\end{subfigure}
\caption{Visualisations of \abbrdef{continuous tempering}{CT} in a bimodal univariate target density. \subref{sfig:bimodal-gm-target-and-gaussian-base} A two-component Gaussian mixture target density (blue curve) and Gaussian base density (green curve) with mean and variance matched to the target. \subref{sfig:jct-energy-and-trajectory} The extended potential energy on the target state $\rvar{x}$ and temperature control variable $\rvar{u}$ (contour plot - dark colours indicate low energy) and an example simulated Hamiltonian trajectory in the joint space (green curve). The temperature control variable bridges the base and target densities lowering energy barriers in the target space. \subref{sfig:jct-prob-dens-and-joint-samples} Joint density on the target state $\rvar{x}$ and inverse temperature $\upbeta$ \eqref{eq:x-beta-joint-density} (contour plot, dark colours indicate high density) and samples from a \abbrref{CT} \abbrref{HMC} chain run in the joint space (circles, size of each circle is proportional to $\pden{\upbeta=1\gvn\rvar{x}=x}$ and so larger symbols indicate a greater weighting in estimates of expectations with respect to the target \eqref{eq:ct-target-expc-estimator}). \subref{sfig:jct-and-hmc-target-histograms} Example target state sample histograms from running standard \abbrref{HMC} in the original target density (top) and running \abbrref{HMC} in the extended joint space (bottom).
}
\label{fig:1d-gm-vis}
\end{figure*}

\section{Hamiltonian Monte Carlo}

\abbrref{HMC} can be used to perform inference whenever the target distribution of interest is defined on a real-valued vector random variable $\rvct{x} \in \reals^D = \set{X}$, by a density function which is differentiable and has support almost everywhere in $\set{X}$. Distributions with bounded support can often be mapped to an equivalent unbounded density by performing a change of variables via a smooth bijective map \citep{carpenter2016stan,kucukelbir2016automatic}.

We will follow the common convention of defining the target density in a Boltzmann-Gibbs form in terms of a real-valued \emph{potential energy} function $\phi : \set{X} \to \reals$ and a (typically unknown) normalising constant $\normconst$ by
\begin{equation}\label{eq:target-density}
  \pden{\rvct{x}=\vct{x}} = \frac{1}{\normconst}\exp\lsb-\phi(\vct{x})\rsb.
\end{equation}

The key idea of \abbrref{HMC} is to simulate a Hamiltonian dynamic in an extended state space to form long-range proposals for a Metropolis accept-reject step with a high probability of acceptance. The \emph{target vector} $\rvct{x}$ is augmented with a \emph{momentum vector} $\rvct{p} \in \reals^D$. Typically the momentum is chosen to be independent of the target state with marginal $\pden{\rvct{p}=\vct{p}} \propto \exp\lsb -\tau(\vct{p})\rsb$ with the joint density factorising as 
\begin{equation}\label{eq:hmc-joint-target}
\begin{split}
\pden{\rvct{x}=\vct{x},\,\rvct{p}=\vct{p}} &= 
\pden{\rvct{x}=\vct{x}}\pden{\rvct{p}=\vct{p}} \\
&\propto 
\exp\lsb -\phi(\vct{x}) - \tau(\vct{p})\rsb.
\end{split}
\end{equation}
With analogy to classical dynamics, $\tau(\vct{x})$ is referred to as the \emph{kinetic energy} and $\hamiltonian(\vct{x},\,\vct{p}) = \phi(\vct{x}) + \tau(\vct{p})$ is termed the \emph{Hamiltonian}. By construction, marginalising the joint density over the momenta recovers $\pden{\rvct{x}=\vct{x}}$.

Each \abbrref{HMC} update involves a discrete-time simulation of the canonical Hamiltonian dynamic 
\begin{equation}\label{eq:standard-hamiltonian-dynamic}
  \td{\vct{x}}{t} = \pd{\hamiltonian}{\vct{p}}\tr = \pd{\tau}{\vct{p}}\tr 
  \qquad 
  \td{\vct{p}}{t} = -\pd{\hamiltonian}{\vct{x}}\tr = -\pd{\phi}{\vct{x}}\tr
\end{equation}
which conserves the Hamiltonian and is time-reversible and volume-preserving. Through choice of an appropriate symplectic integrator such as the popular leapfrog (St\"ormer-Verlet) scheme, the simulated discrete time dynamic remains exactly time-reversible and volume-preserving and also approximately conserves the Hamiltonian even over long simulated trajectories \citep{leimkuhler2004simulating}.

The discrete-time dynamic is used to generate a new proposed state $(\vct{x}',\,\vct{p}')$ given the current state $(\vct{x},\,\vct{p})$ and this proposal accepted or rejected in a Metropolis step with acceptance probability $\min\lbr 1,\,\exp\lsb \hamiltonian(\vct{x},\,\vct{p})- \hamiltonian(\vct{x}',\,\vct{p}')\rsb\rbr$. Due to the approximate Hamiltonian conservation the acceptance probability is typically close to one, and so \abbrref{HMC} is able to make long-range moves in high-dimensional state-spaces while still maintaining high acceptance rates, a significant improvement over the behaviour typical of simpler Metropolis--Hastings methods in high-dimensions.

The energy conservation property which gives this desirable behaviour also however suggests that standard \abbrref{HMC} updates are unlikely to move between isolated modes in a target distribution. The Hamiltonian is approximately conserved over a trajectory therefore $\phi(\vct{x}') - \phi(\vct{x}) \approx \tau(\vct{p}) - \tau(\vct{p}')$. Typically a quadratic kinetic energy  $\tau(\vct{p}) = \vct{p}\tr\mtx{M}^{-1}\vct{p} \,/\,2$ is used corresponding to a Gaussian marginal density on the momentum. As this kinetic energy is bounded below by zero, the maximum change in potential energy over a trajectory is approximately equal to the initial kinetic energy. 

At equilibrium the momenta will have a Gaussian distribution and so the kinetic energy a $\chi^2$ distribution with mean $\nicefrac{D}{2}$ and variance $D$ \citep{neal2011mcmc,betancourt2013general}. If potential energy barriers significantly larger than $\sim D$ separate regions of the configuration state space the \abbrref{HMC} updates are unlikely to move across the barriers meaning impractically long sampling runs will be needed for effective ergodicity.

\section{Thermodynamic methods}

A common approach in \abbrref{MCMC} methods for dealing with multimodal target distributions is to introduce a concept of temperature. In statistical mechanics, the Boltzmann distribution on a configuration $\rvct{x}$ of a mechanical system with energy function $\phi$ and in thermal equilibrium with a heat bath at temperature $T$ is defined by a probability density $\exp\lsb-\beta\phi(\vct{x})\rsb / \partfunc(\beta)$ where $\beta = (k_B T)^{-1}$ is the \emph{inverse temperature}, $k_B$ is Boltzmann's constant and $\partfunc(\beta)$ is the \emph{partition function}. At high temperatures ($\beta \to 0$) the density function becomes increasingly flat across the target state space and, correspondingly, energy barriers between different regions of the state space become lower.

In the above statistical mechanics formulation, if $\rvct{x} \in \reals^D$ the distribution in the limit $\beta \to 0$ would be an improper flat density across the target state space. More usefully from a statistical perspective we can use an inverse temperature variable $\beta \in [0, 1]$ to geometrically bridge between a simple \emph{base distribution} with normalised density $\exp\lsb-\psi(\vct{x})\rsb$ at $\beta=0$ and the target distribution at $\beta=1$
\begin{align}
  \pi\lpa\vct{x} \gvn \beta \rpa &=
  \exp\lsb -\beta\phi(\vct{x}) - (1-\beta)\psi(\vct{x}) \rsb, \label{eq:geometric-bridge} \\
  \partfunc(\beta) &= 
  \int_{\set{X}} 
    \exp\lsb -\beta\phi(\vct{x}) - (1-\beta)\psi(\vct{x}) \rsb
  \,\dr\vct{x}. \label{eq:partition-function}
\end{align}
Several approaches for varying the system temperature in \abbrref{MCMC} methods have been proposed. In \emph{simulated tempering} \citep{marinari1992simulated} an ordered set of inverse temperatures are chosen
\begin{equation}\label{eq:finite-inv-temp-schedule}
  \fset{\beta_n}_{n=0}^N : 0 = \beta_0 < \beta_1 < \dots < \beta_N = 1,
\end{equation}
and a joint distribution defined as
\begin{equation}\label{eq:simulated-tempering-joint}
  \pden{\rvct{x}=\vct{x},\,\upbeta = \beta_n} \propto
  \pi\lpa\vct{x} \gvn \beta_n \rpa \exp\lpa w_n\rpa,
\end{equation}
where $\lbrace w_n\rbrace_{n=0}^N$ are a set of prior weights associated with each inverse temperature. Alternating \abbrref{MCMC} updates of $\upbeta$ given $\rvct{x}$ and $\rvct{x}$ given $\upbeta$ are performed, with $\rvct{x}$ samples for which $\upbeta=1$ converging in distribution to the target. Updates of the inverse temperature variable can propose moves to a limited set of neighbouring inverse temperatures or sample $\upbeta$ independently from its conditional $\prob{\upbeta=\beta_n\gvn\rvct{x}=\vct{x}}$.

The ratio of the marginal probabilities of $\upbeta=1$ and $\upbeta=0$ can be related to the usually unknown normalising constant $\normconst$ for the target distribution by
\begin{equation}\label{eq:st-marginal-ratio}
\normconst = \exp(w_0 - w_N) \, \frac{\prob{\upbeta=1}}{\prob{\upbeta=0}},
\end{equation}
allowing estimation of $\normconst$ from a simulated tempering chain by computing the ratio of counts of $\upbeta=1$ and $\upbeta=0$.

As an alternative \cite{carlson2016partition} proposes to instead use a `\emph{Rao-Blackwellised}' estimator for $\normconst$ based on the identity
\begin{equation}\label{eq:rao-blackwell-tempered-sampling}
  \prob{\upbeta = \beta_n} = 
  \int_{\set{X}} 
    \prob{\upbeta=\beta_n\gvn\rvct{x}=\vct{x}} 
    \pden{\rvct{x}=\vct{x}} 
  \,\dr\vct{x} 
\end{equation}
which indicates $\prob{\upbeta = \beta_n}$ can be estimated by averaging the conditional probabilities $\prob{\upbeta=\beta_n\gvn\rvct{x}=\vct{x}}$ across samples of $\rvct{x}$ from the joint in \eqref{eq:simulated-tempering-joint}. The authors empirically demonstrate this estimator can give significantly improved estimation accuracy over the simpler count-based estimator.

A problem for simulated tempering methods is that as  $\prob{\upbeta = \beta_n} \propto \partfunc(\beta_n) \,\exp( w_n)$ and the partition function can vary across several orders of magnitude, the chain can mix poorly between inverse temperatures. This is often tackled with an iterative approach in which initial pilot runs are used to estimate $\prob{\upbeta = \beta_n}$ and the weights $\fset{w_n}_{n=0}^N$ set so as to try to flatten out this marginal distribution \cite{geyer1995annealing}.

An alternative is \emph{parallel tempering}, \emph{Metropolis-coupled MCMC} or \emph{replica exchange} \citep{swendsen1986replica,geyer1991markov,earl2005parallel}, where multiple Markov chains on the target state are run in parallel, with the $n$\textsuperscript{th} chain having an associated inverse temperature $\beta_n$ defined as in \eqref{eq:finite-inv-temp-schedule}. \abbrref{MCMC} updates are performed on each chain which leave $\pi(\vct{x} \gvn \beta_n)$ invariant. Interleaved with these updates, exchanges of the configuration states of the chains at adjacent inverse temperatures $(\beta_n,\,\beta_{n+1})$ are proposed and accepted or rejected in a Metropolis--Hastings step. Parallel tempering sidesteps the issue with mixing between inverse temperatures by keeping the inverse temperatures for the chains fixed, at a cost of a significantly increased state size.

\abbrdef{Annealed Importance Sampling}{AIS} \citep{neal2001annealed} is another thermodynamic ensemble method, closely related to \emph{Tempered Transitions} \citep{neal1996sampling} and often the `go-to' method for normalising constant estimation in the machine learning literature e.g \citep{salakhutdinov2008quantitative,wu2016quantitative}. In \abbrref{AIS} an ordered set of inverse temperatures are defined as in \eqref{eq:finite-inv-temp-schedule} and a corresponding series of transition operators $\lbrace \mathbb{T}_n \rbrace_{n=1}^{N-1}$ which have $\pi(\vct{x}\gvn\beta_n)$ for the corresponding $n$ as their stationary distributions. Assuming the base distribution corresponding to $\beta_0 = 0$ can be sampled from independently, an independent state $\vct{x}^{(0)}$ from this base distribution is generated and then the transition operators applied in a fixed sequence $\transition_{1} \dots \transition_{N-1}$ to generate a chain of intermediate states $\lbrace \vct{x}^{(n)} \rbrace_{n=1}^{N-1}$. The product of ratios
\begin{equation}\label{eq:ais-weights}
  r = \prod_{n=0}^{N-1} \lsb \frac{\pi\lpa\vct{x}^{(n)}\gvn\beta_{n+1}\rpa}{\pi\lpa\vct{x}^{(n)}\gvn\beta_n\rpa} \rsb
\end{equation}
is an importance weight for the final state $\vct{x}^{(N-1)}$ with respect to the target distribution \eqref{eq:target-density} and also an unbiased estimator for $\normconst$. By simulating multiple independent runs of this process, a set of importance weighted samples can be computed to estimate expectations with respect to the target \eqref{eq:target-density} and the weights $r$ used to unbiasedly estimate $\normconst$. Due to Jensen's inequality the unbiased estimate of $\normconst$ corresponds to stochastic lower bound on $\log \normconst$ \citep{grosse2015sandwiching}.

In cases where an exact sample from the target distribution can be generated, running \emph{reverse \abbrref{AIS}} from the target to base distribution can be used to calculate an unbiased estimate of $1 / \normconst$ and so a stochastic upper bound on $\log \normconst$ \citep{grosse2015sandwiching}. This combination of generating a stochastic lower bound on $\log \normconst$ from forward \abbrref{AIS} runs and a stochastic upper bound from reverse \abbrref{AIS} runs is termed \emph{bidirectional Monte Carlo}.


\section{Using a continuous temperature}

In all three of \abbrref{AIS}, parallel and simulated tempering the choice of the discrete inverse temperature schedule \eqref{eq:finite-inv-temp-schedule} is key to getting the methods to perform well in complex high dimensional distributions \citep{neal1996sampling,behrens2012tuning,betancourt2014adiabatic}. To get reasonable performance it may be necessary to do preliminary pilot runs to guide the number of inverse temperatures and spacing between them to use, adding to the computational burden and difficulty of using these methods in a black-box fashion. A natural question is therefore whether it is possible to use a continuously varying inverse temperature variable to side-step the need to choose a set of values. 

\emph{Path sampling} \citep{gelman1998simulating} proposes this approach, defining a general \emph{path} as a function parametrised by $\beta$ which continuously maps between the target density $\exp\lsb-\phi(\vct{x})\rsb / \normconst$  at $\beta=1$ and a base density $\exp\lsb-\psi(\vct{x})\rsb$ at $\beta = 0$, with the geometric bridge in \eqref{eq:geometric-bridge} a particular example. A joint target $\pden{\rvct{x}=\vct{x},\,\upbeta=\beta} \propto \pi(\vct{x}\gvn\beta) \,\rho(\beta)$ is defined with $\rho(\beta)$ a user-chosen prior density on the inverse temperature variable analogous to the weights $\lbr w_n \rbr_{n=0}^N$ in simulated tempering. 

In \citep{gelman1998simulating} it is proposed to construct a Markov chain leaving the joint density invariant by alternating updates of $\rvct{x}$ given $\upbeta$ and $\upbeta$ given $\rvct{x}$. Samples from the joint system can be used to estimate $\normconst$ via the \emph{thermodynamic integration} identity
\begin{equation*}\label{thermodynamic-integration}
\log \normconst = 
\int_0^1 \hspace{-0.5em} \int_{\set{X}}
  \frac{\pden{\rvct{x}=\vct{x},\,\upbeta=\beta}}{\pden{\upbeta=\beta}}\,
  \pd{\log\pi(\vct{x}\gvn\beta)}{\beta}
\,\dr\vct{x}\,\dr\beta.
\end{equation*}

\emph{Adiabatic Monte Carlo} \citep{betancourt2014adiabatic} also proposes using a continuously varying inverse temperature variable, here specifically in the context of \abbrref{HMC}. The original Hamiltonian system $(\rvct{x},\,\rvct{p})$ is further augmented with a continuous inverse temperature coordinate $\upbeta \in [0,\,1]$. 
A \emph{contact Hamiltonian} is defined on the augmented system, 
\begin{equation}\label{eq:contact-hamiltonian}
\begin{split}
  \hamiltonian_c(\vct{x},\,\vct{p},\,\beta) = \beta \phi(\vct{x}) &+ ( 1 - \beta ) \psi(\vct{x}) 
  + \frac{1}{2}\vct{p}\tr\mtx{M}^{-1}\vct{p} \\
  &+ \log \partfunc(\beta) + \hamiltonian_0
\end{split}
\end{equation}
this defining a corresponding \emph{contact Hamiltonian flow}, 
\begin{equation}\label{eq:contact-hamiltonian-flow}
  \td{\vct{x}}{t} = \pd{\hamiltonian_c}{\vct{p}}\tr, ~
  \td{\vct{p}}{t} = \pd{\hamiltonian_c}{\beta} \vct{p} -\pd{\hamiltonian_c}{\vct{x}}\tr, ~
  \td{\beta}{t} = \hamiltonian_c - \pd{\hamiltonian_c}{\vct{p}}\vct{p}.
\end{equation}
The contact Hamiltonian flow restricted to the zero level-set of the contact Hamiltonian (which the initial state can always be arranged to lie in by appropriately choosing the arbitrary constant $\hamiltonian_0$) generates trajectories which exactly conserve the contact Hamiltonian and extended state space volume element, and correspond to the thermodynamical concept of an isentropic or reversible adiabatic process. 

For a quadratic kinetic energy, $\td{\beta}{t}$ is always non-positive and so forward simulation of the contact Hamiltonian flow generates non-increasing trajectories in $\beta$ (and backwards simulation generates non-decreasing trajectories in $\beta$). In the ideal case this allows the inverse temperature range $[0,\,1]$ to be coherently traversed without the random walk exploration inherent to simulated tempering.

Simulating the contact Hamiltonian flow is non-trivial in practice however: the contact Hamiltonian \eqref{eq:contact-hamiltonian} depends on the log partition function $\log \partfunc(\beta)$, the partial derivatives of which require computing expectations with respect to $\pi(\vct{x}\gvn\beta)$ which for most problems is intractable to do exactly. Moreover the contact flow can encounter meta-stabilities whereby $\td{\beta}{t}$ becomes zero and the flow halts at an intermediate $\beta$ meaning the flow no longer defines a bijection between $\beta=0$ and $\beta=1$. This can be ameliorated by regular resampling of the momenta however this potentially increases the random-walk behaviour of the overall dynamic.

An alternative \emph{extended Hamiltonian approach} for simulating a system with a continuously varying inverse temperature was proposed recently in the statistical physics literature \citep{gobbo2015extended}. The inverse temperature of the system is indirectly set via an auxiliary variable, which we will term a \emph{temperature control variable} $\rvar{u} \in \reals$. This control variable is mapped to an interval $[s,\,1]$, $0 < s < 1$ via a smooth piecewise defined function $\beta : \reals \to [s,\,1]$, with the conditions that for a pair of thresholds $(\theta_1,\,\theta_2)$ with $0 < \theta_1 < \theta_2$, $\beta(u) = 1 ~\forall~ |u| \leq \theta_1$, $\beta(u) = s ~\forall~ |u| \geq \theta_2$ and $s < \beta(u) < 1 ~\forall~ \theta_1 < |u| < \theta_2$.

Unlike Adiabatic Monte Carlo, an additional momentum variable $\rvar{v}$ corresponding to $\rvar{u}$ is also introduced. Although seemingly a minor difference this simplifies the implementation of the approach significantly as the system retains a symplectic structure and can continue to be viewed within the usual Hamiltonian dynamics framework. An \emph{extended Hamiltonian} is then defined on the augmented system
\begin{equation}\label{eq:extended-hamiltonian-gobbo-leimkuhler}
  \extended{\hamiltonian}(\vct{x},\,u,\,\vct{p},\,v) = 
  \beta(u) \phi(\vct{x}) + \omega(u) + \frac{1}{2} \vct{p}\tr\mtx{M}^{-1}\vct{p} + \frac{v^2}{2m}
\end{equation}
where $\omega$ is a `confining potential' on $u$ and $m$ is the mass (marginal variance) associated with $\rvar{v}$. 

This extended Hamiltonian remains separable with respect to the extended configuration $(\rvct{x},\,\rvar{u})$ and extended momentum $(\rvct{p},\,\rvar{v})$ and so can be efficiently simulated using, for example, a standard leapfrog integrator. In \cite{gobbo2015extended} the extended Hamiltonian dynamics are integrated using a Langevin scheme without Metropolis adjustment and shown to improve mixing in several molecular dynamics problems. 

Due to the condition $\beta(u) = 1 ~\forall~ |u| < \theta_1$, the set of sampled configuration states $\rvct{x}$ which have associated $|\rvar{u}| < \theta_1$ will (assuming the dynamic is ergodic and Metropolis adjustment were used) asymptotically converge in distribution to the target, and so can be used to estimate expectations without any importance re-weighting.

The $\beta$ function is required to be bounded below by some $s > 0$ in \citep{gobbo2015extended} due to the base density being bridged to being an improper uniform density across $\set{X}$. The partition function $\partfunc(\beta) \to \infty$ as $\beta \to  0$ in this case, which would imply an infinite density for regions in the extended state space where $\beta(\rvar{u}) = 0$. Even with a non-zero lower bound on $\beta$, the large variations in $\partfunc\lsb\beta(\rvar{u})\rsb$  across different $\rvar{u}$ values can lead to the dynamic poorly exploring the $\rvar{u}$ dimension. 

In \citep{gobbo2015extended} this issue is tackled by introducing an adaptive history-dependent biasing potential on $\rvar{u}$ to try to achieve a flat density across a bounded interval $|\rvar{u}| < \theta_2$, using for example metadynamics \citep{laio2002escaping}. The resulting non-Markovian updates bias the invariant distribution of the target state however this can be accounted for either by a re-weighting scheme \citep{bonomi2009reconstructing}, or using a vanishing adaptation.

\section{Proposed approach}\label{sec:proposed-approach}

As in \citep{gobbo2015extended} we define an extended Hamiltonian on an augmented state $(\rvct{x},\,\rvar{u})$ with associated momenta $(\rvct{p},\,\rvar{v})$,
\begin{equation}\label{eq:extended-hamiltonian}
\begin{split}
  \extended{\hamiltonian}(\vct{x},\,u,\,\vct{p},\,v) =
  \beta(u) \lsb \phi(\vct{x}) + \log\zeta\rsb + \\
  \lsb 1 - \beta(u) \rsb \psi(\vct{x}) - \log\left|\pd{\beta}{u}\right| +
  \frac{1}{2} \vct{p}\tr\mtx{M}^{-1}\vct{p} + \frac{v^2}{2m}.
\end{split}
\end{equation}

Like the previous extended Hamiltonian approach of \citep{gobbo2015extended}, this Hamiltonian is separable and the corresponding dynamic can be efficiently simulated with a leapfrog integrator. The reversible and volume-preserving simulated dynamic can then be used as a proposal generating mechanism on the joint space $(\rvct{x},\,\rvar{u},\,\rvct{p},\,\rvar{v})$ for a Metropolis--Hastings step as in standard \abbrref{HMC}. We will term this approach of running \abbrref{HMC} in the extended joint space \emph{joint continuous tempering}.

In contrast to \citep{gobbo2015extended} we propose to use a smooth monotonically increasing map $\beta : \reals \to [0, 1]$ as the inverse temperature control function, with our default choice in all experiments being the logistic sigmoid $\beta(u) = \lsb 1 + \exp(-u)\rsb^{-1}$.

As in the discussion earlier $\psi$ is the negative logarithm of a simple \emph{normalised} base density, which (as we will motivate in the next section) we will usually choose to be an approximation to the target density \eqref{eq:target-density}. Similarly the $\log\zeta$ term will be chosen to be an approximation to $\log \normconst$.

We can marginalise out the momenta from the joint distribution defined by this Hamiltonian, giving a joint density
\begin{equation}\label{eq:x-u-joint-density}
\begin{split}
\pden{\rvct{x}=\vct{x},\,\rvar{u}=u} \propto \\
\left|\pd{\beta}{u}\right|
\exp\lsb
  -\beta(u)\lpa \phi(\vct{x}) + \log\zeta\rpa
  - \lpa 1 - \beta(u)\rpa\psi(\vct{x})
\rsb.
\end{split}
\end{equation}
If we define a random variable $\upbeta = \beta(\rvar{u})$ and use the change of variables formula for a density, we further have that
\begin{equation}\label{eq:x-beta-joint-density}
\pden{\rvct{x}=\vct{x},\,\upbeta=\beta} \propto \frac{
\exp \lsb
  -\beta \phi(\vct{x}) - ( 1 - \beta)\psi(\vct{x})
\rsb}{\zeta^\beta}.
\end{equation}

The bijectivity between $\rvar{u}$ and $\upbeta$ is useful as although if simulating a Hamiltonian dynamic we will generally wish to work with $\rvar{u}$ as it is unbounded, the conditional density on $\upbeta$ given $\rvct{x}$ has a tractable normalised form
\begin{equation}\label{eq:beta-gvn-x-density}
\pden{\upbeta=\beta \gvn\rvar{x}=\vct{x}} = 
\frac{\exp\lsb - \beta\Delta(\vct{x})\rsb\Delta(\vct{x})}{1 - \exp\lsb-\Delta(\vct{x})\rsb},
\end{equation}
with $\Delta(\vct{x}) = \phi(\vct{x}) + \log\zeta - \psi(\vct{x})$. This corresponds to an exponential distribution with rate parameter $\Delta(\vct{x})$ truncated to $[0, 1]$. As an alternative to the joint updates, another option is therefore to form a Markov chain which leaves \eqref{eq:x-beta-joint-density} invariant by alternating independently sampling $\upbeta$ from its conditional given $\rvct{x}$ and performing a transition which leaves the conditional on $\rvct{x}$ given $\upbeta$ invariant, similar to the suggested approach in \citep{gelman1998simulating}. We will term this Gibbs sampling type procedure as \emph{Gibbs continuous tempering}.

Further we can use \eqref{eq:beta-gvn-x-density} to write the marginal density on $\upbeta$
\begin{equation}\label{eq:ct-beta-marginal}
\pden{\upbeta=\beta} =
\expc{\frac{\exp\lsb - \beta\Delta(\rvct{x})\rsb\Delta(\rvct{x})}{1 - \exp\lsb-\Delta(\rvct{x})\rsb}}.
\end{equation}

By integrating the joint \eqref{eq:x-beta-joint-density} over $\set{X}$ we also have that
\begin{equation}\label{eq:beta-marginal-0-1}
\begin{split}
\pden{\upbeta=\beta} = \frac{1}{C\zeta^\beta} \int_{\set{X}} 
  \pi(\vct{x}\gvn\beta)
\,\dr\vct{x} =
\frac{\partfunc(\beta)}{C\zeta^\beta}\\
\Rightarrow ~
\pden{\upbeta=0} = \frac{1}{C}
,~
\pden{\upbeta=1} = \frac{\normconst}{C \zeta},
\end{split}
\end{equation}
for an unknown normalising constant $C$ of the joint, and so for \abbrref{MCMC} samples $\lbr \vct{x}^{(s)},\,\beta^{(s)}\rbr_{s=1}^S$ from the joint \eqref{eq:x-beta-joint-density}
\begin{equation}\label{eq:ct-norm-const-estimator}
\normconst = \frac{\pden{\upbeta=1}}{\pden{\upbeta=0}} \zeta = 
\lim_{S\to\infty} \frac
{\sum_{s=1}^S\lsb w_1\lpa \vct{x}^{(s)}\rpa\rsb}
{\sum_{s=1}^S\lsb w_0\lpa \vct{x}^{(s)}\rpa\rsb}
\zeta,
\end{equation}
\begin{equation*}
\textrm{with }
w_0(\vct{x}) = \frac{\Delta(\vct{x})}{1 - \exp\lsb-\Delta(\vct{x})\rsb},~
w_1(\vct{x}) = \frac{\Delta(\vct{x})}{\exp\lsb\Delta(\vct{x})\rsb - 1}.
\end{equation*}
This can be seen to be a continuous analogue of the \emph{Rao-Blackwellised} estimator combining \eqref{eq:st-marginal-ratio} and \eqref{eq:rao-blackwell-tempered-sampling} used in \cite{carlson2016partition}.

Similarly we can calculate consistent estimates of expectations with respect to the target density $\pden{\rvct{x}=\vct{x}\gvn\upbeta=1} = \exp\lsb-\phi(\vct{x})\rsb / Z$ as importance weighted sums
\begin{align}\label{eq:ct-target-expc-estimator}
\expc{f(\rvct{x})\gvn\upbeta=1} &=
\frac{
  \int_{\set{X}} 
    f(\vct{x})\pden{\upbeta=1 \gvn\rvct{x}=\vct{x}} \pden{\rvct{x} = \vct{x}} 
  \,\dr\vct{x}
}{
  \int_{\set{X}} 
    \pden{\upbeta=1 \gvn\rvct{x}=\vct{x}} \pden{\rvct{x} = \vct{x}} 
  \,\dr\vct{x}
} 
\nonumber\\
&=
\lim_{S\to\infty} 
\frac
{\sum_{s=1}^S \lsb w_1\lpa \vct{x}^{(s)}\rpa f\lpa\vct{x}^{(s)}\rpa\rsb}
{\sum_{s=1}^S\lsb w_1\lpa \vct{x}^{(s)}\rpa \rsb}.
\end{align}
We can also estimate expectations with respect to the base density $\pden{\rvct{x}=\vct{x}\gvn\upbeta=0} = \exp\lsb-\psi(\vct{x})\rsb$ using
\begin{equation}\label{eq:ct-base-expc-estimator}
\expc{f(\rvct{x})\gvn\upbeta=0}
=
\lim_{S\to\infty} 
\frac
{\sum_{s=1}^S \lsb w_0\lpa \vct{x}^{(s)}\rpa f\lpa\vct{x}^{(s)}\rpa\rsb}
{\sum_{s=1}^S\lsb w_0\lpa \vct{x}^{(s)}\rpa \rsb}.
\end{equation}
Often the base density will have known moments (e.g. mean and covariance of a Gaussian base density) which can be compared to the estimates calculated using \eqref{eq:ct-base-expc-estimator} to check for convergence problems. Convergence of the estimates to the true moments is not a sufficient condition for convergence of the chain to the target joint density \eqref{eq:x-beta-joint-density} but is necessary.

\section{Choosing a base density}\label{sec:choosing-base-density}

By applying H\"older's and Jensen's inequalities we can bound $\pden{\upbeta=\beta}$ (see Appendix A for details)
\begin{equation}\label{eq:partition-function-bounds}
  \frac{1}{C}\lpa\frac{\normconst}{\zeta}\rpa^\beta \exp\lpa -\beta d^{b\to t}\rpa
  \leq \pden{\upbeta=\beta} \leq
  \frac{1}{C}\lpa\frac{\normconst}{\zeta}\rpa^\beta,
\end{equation}
where $d^{b\to t}$ indicates the \abbrdef{Kullback--Leibler}{KL} divergence from the base to target distribution
\begin{equation}\label{eq:kl-base-to-target}
  d^{b\to t} = \int_{\set{X}} 
    \exp\lsb -\psi(\vct{x})\rsb \log\lpa\frac{\exp\lsb -\psi(\vct{x})\rsb}{\exp\lsb -\phi(\vct{x})\rsb / \normconst}\rpa
  \,\dr\vct{x}.
\end{equation} 
If $\zeta = \normconst$ the upper-bound is constant. If additionally  we had $d^{b\to t} = 0$, the bound becomes tight and we would have a flat marginal density on $\upbeta$, which although is not guaranteed to be optimal we hope will improve mixing in the $\upbeta$ dimension.

In reality we do not know $\normconst$ and cannot choose a base distribution such that the KL divergence is zero as we wish to use a simple density amenable to exploration. However we can see that under the constraint of the base distribution allowing exploration, a potentially useful heuristic is to minimise the KL divergence to the target distribution. Further we want to find a $\zeta$ as close to $\normconst$ as possible. 

Variational inference is an obvious route for tackling both problems, allowing us to fit a base density in a simple parametric family (e.g. Gaussian) by directly minimising the KL divergence \eqref{eq:kl-base-to-target} while also giving a lower bound on $\log \normconst$. In some cases we can use variational methods specifically aimed at the target distribution family. More generally methods such as \abbrdef{Automatic Differentiation Variational Inference}{ADVI} \citep{kucukelbir2016automatic} provide a black-box framework for fitting variational approximations to differentiable target densities. 

In some models such as Variational Autoencoders \citep{kingma2013auto,rezende2014stochastic} a parametric variational approximation to the target density of interest (e.g. posterior on latent space) is fitted during training of the original model, in which case it provides a natural choice for the base distribution as observed in \citep{wu2016quantitative}. 

A potential problem is that the classes of target distribution that we are particularly interested in applying our approach to --- those with multiple isolated modes --- are precisely the same distributions that simple variational approximations will tend to fit poorly, the divergence \eqref{eq:kl-base-to-target} being minimised favouring `mode-seeking' solutions \citep{bishop2006pattern}, which usually fit only one mode well. This both limits how small the divergence from the base to target can be made, but also crucially is undesirable as we wish the base distribution to allow the dynamic to move between modes in the target.

One option would be to instead to minimise the reversed form of the KL divergence from the target to base distribution, this tending to produce `mode-covering' solutions that match global moments (and can also be used to produce an analogous lower bound on the marginal density to that for $d^{b\to t}$ in \eqref{eq:partition-function-bounds} as shown in Appendix A). Methods such as \abbrdef{expectation propagation}{EP} \citep{minka2001expectation} do allow moment-matching approximations to be found and may be a good option in some cases. Another possibility would be to use an alternative divergence in a variational framework that favours mode-covering solutions, with methods using the $\chi$-divergence \citep{dieng2016chi} and R\'enyi divergence \citep{li2016renyi} having been recently proposed in this context.

A further alternative is to fit multiple local variational approximations $\lbrace q_i(\vct{x}) \rbrace_{i=1}^L$ by minimising the variational objective from multiple random parameter initialisations (discarding any duplicate solutions as measured by some distance tolerance between the variational parameters), each approximating a single mode well. We can then combine these local approximations into a global approximation $q(\vct{x})$. 

One option is to use a mixture of the local approximations 
\begin{equation}\label{eq:variational-mixture-approx}\textstyle
  q(\vct{x}) = \frac{1}{\zeta} \sum_{i=1}^L \lsb \exp(\ell_i) q_i(\vct{x}) \rsb,\quad
  \zeta = \sum_{i=1}^L \exp(\ell_i),
\end{equation}
with $\ell_i$ the final variational objective value for $q_i$ ($\log \normconst$ minus the KL divergence from $q_i$ to target distribution). If the local approximations have non-overlapping support this will lead to a global approximation which is guaranteed to be at least as close in \abbrref{KL} divergence as any of the local approximations and a $\log \zeta$ which is at least as tight a lower bound on $\log Z$ as any of the individual $\ell_i$ \citep{zobay2009mean}. 

Often we may wish to use local approximations with overlapping support (e.g. Gaussian) where the guarantee does not apply. However for the cases of target distributions with multiple isolated modes the `overlap' (regions with high density under multiple $q_i$) between local Gaussian approximations to each mode will often be minimal and so the method is still a useful heuristic.

A mixture distribution is unlikely to itself be a good choice of base distribution however, as it will tend to be multimodal. We therefore instead propose to use a base distribution with moments matched to the fitted mixture distribution , e.g. a Gaussian $\exp[-\psi(\vct{x})]$ with mean and covariance matched to the mean and covariance of the mixture $q(\vct{x})$.

For complex target distributions it will sometimes remain difficult to find a reasonable $\psi$ and $\log \zeta$ which well approximate $\phi$ and $\log \normconst$ using approaches such as those just discussed . If the \abbrref{KL} divergence from the base to target distribution is high, the bounds on the marginal in \eqref{eq:partition-function-bounds} become increasingly loose and in practice this tends to lead to low densities on intermediate inverse temperatures. This reduces the ability of the \abbrref{MCMC} dynamic to move between the inverse temperatures corresponding to the target and base distributions and so the mixing gain from the augmentation. 

Similarly from \eqref{eq:beta-marginal-0-1} the density ratio $\pden{\upbeta = 1} \,/ \pden{\upbeta = 0}$ is $\exp(\log \normconst - \log \zeta)$, and so for large $|\log\normconst - \log\zeta|$ the dynamic will tend to spend most of the time close to $\upbeta = 1$ if $\log \normconst  > \log \zeta$ or $\upbeta = 0$ if $\log \normconst < \log \zeta$. In the former case there will be limited gain from using the extended space over running a chain on the original target density while in the latter the variance of the estimator in \eqref{eq:ct-target-expc-estimator} will be comparable to directly using an importance sampling estimator for the target using samples from the base density.

A natural approach to try to ameliorate these effects is to use an iterative `bootstrap' method. An initial $\psi$ and $\log \zeta$ are chosen for example using a Gaussian variational approximation to the target density as described above. A Markov chain which leaves the joint density \eqref{eq:x-beta-joint-density} invariant is then simulated. The samples from this chain can then be used to both compute an estimate for $\log \normconst$ using \eqref{eq:ct-norm-const-estimator} and updated estimates of the target density mean and covariance using \eqref{eq:ct-target-expc-estimator}. A new Gaussian base density can then be specified with the updated mean and covariance estimates and $\log \zeta$ chosen to be the updated $\log \normconst$ estimate. Samples of the new joint density can then be used to update $\log \zeta$ and $\psi$ again and so on. This is analogous to the iterative approaches often used in simulated tempering to choose the prior weights on the inverse temperature values \citep{geyer1995annealing,carlson2016partition}. 



\section{Experiments}\label{sec:experiments}

To validate the proposed approach we performed a series of experiments comparing our proposed continuous tempering methods to existing approaches. Code for running all experiments is available at \url{https://git.io/cthmc}.

\subsection{One-dimensional bimodal target}\label{subsec:exp-bimodal-target}

We start with an illustrative example of the gains of the proposed approach over standard \abbrref{HMC} in target densities with isolated modes. We use a one-dimensional Gaussian mixture density with two separated Gaussian components as the target density, shown by the blue curve in Figure \ref{sfig:bimodal-gm-target-and-gaussian-base}. Although performance in this toy univariate model is not necessarily reflective of that in more realistic higher-dimensional models, it has the advantage of allowing the joint density on $\rvar{x}$ and  $\upbeta = \beta(\rvar{u})$ to be directly visualised.

For the base density $\exp[-\psi(x)]$ we use a univariate Gaussian with mean and variance matched to those of the target density (corresponding to the Gaussian density minimising the \abbrref{KL} divergence from the target to base distribution), shown by the green curve in Figure \ref{sfig:bimodal-gm-target-and-gaussian-base}. We also set $\log \zeta = \log Z$ and so the performance here represents a `best-case' scenario for the continuous tempering approach.

The resulting potential energy ($-\log\pden{\rvar{x},\,\rvar{u}}$) on the extended $(\rvar{x},\,\rvar{u})$ space is shown in Figure \ref{sfig:jct-energy-and-trajectory}. For positive temperature control values (and so inverse temperature values close to 1), the energy surface tends increasingly to the double-well potential corresponding to the target distribution, with a high energy barrier between the two modes. For negative temperature control values the energy surface tends towards the single quadratic well corresponding to the Gaussian base density. The resulting joint energy surface allows for paths between the values of the target state $\rvar{x}$ corresponding to the two modes which have much lower potential energy barriers than the potential barrier between the two modes in the original target space, allowing simulated Hamiltonian trajectories such as that shown in green to more easily explore the target state space. 

Samples from a \abbrref{HMC} chain on the extended joint space are shown in Figure \ref{sfig:jct-prob-dens-and-joint-samples}, with the joint density on $(\rvar{x},\,\upbeta)$ \eqref{eq:x-beta-joint-density} shown in the background as a contoured heat map. It can be seen that the Hamiltonian dynamic is able to explore the joint space well with good coverage of all of the high density regions. The size of the points in \ref{sfig:jct-prob-dens-and-joint-samples} is proportional to $w_1(x) = \pden{\upbeta=1\gvn\rvar{x}=x}$ and so reflects the importance weights of the samples in the estimator for expectations with respect to the target in \eqref{eq:ct-target-expc-estimator}. Importantly even points for which $\upbeta$ is close to zero can contribute significantly to the expectations if the corresponding $\rvar{x}$ value is probable under the target: this is in contrast to the extended Hamiltonian approach of \citep{gobbo2015extended} where only a subset of points corresponding to $\upbeta=1$ are used to compute expectations.

The final panel, Figure \ref{sfig:jct-and-hmc-target-histograms} shows empirical histograms on the target variable $\rvar{x}$ estimated from samples of a chain on the extended space (joint continuous tempering, bottom) and standard \abbrref{HMC} on the original target space (top). As can be seen the standard \abbrref{HMC} approach gets stuck in one mode thus does not assign any mass to the other mode in the histogram, unlike the tempered chain which identifies both modes and accurately estimates their relative masses.

\subsection{Gaussian mixture Boltzmann machine relaxations}\label{subsec:exp-bm-relaxations}

\begin{figure*}[!ht]
\centering
\begin{subfigure}[b]{.33\linewidth}
\vskip 0pt
\centering
\includegraphics[width=0.95\textwidth]{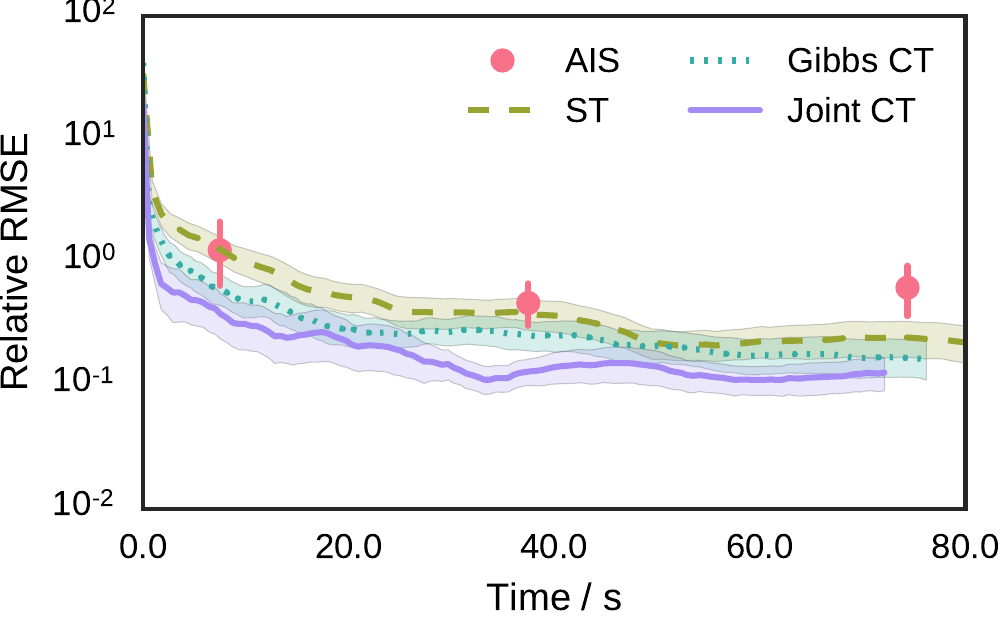} 
\caption{$\log \normconst$}\label{sfig:bmr-30-unit-scale-6-log-norm}
\end{subfigure}%
\begin{subfigure}[b]{.33\linewidth}
\vskip 0pt
\centering
\includegraphics[width=0.95\textwidth]{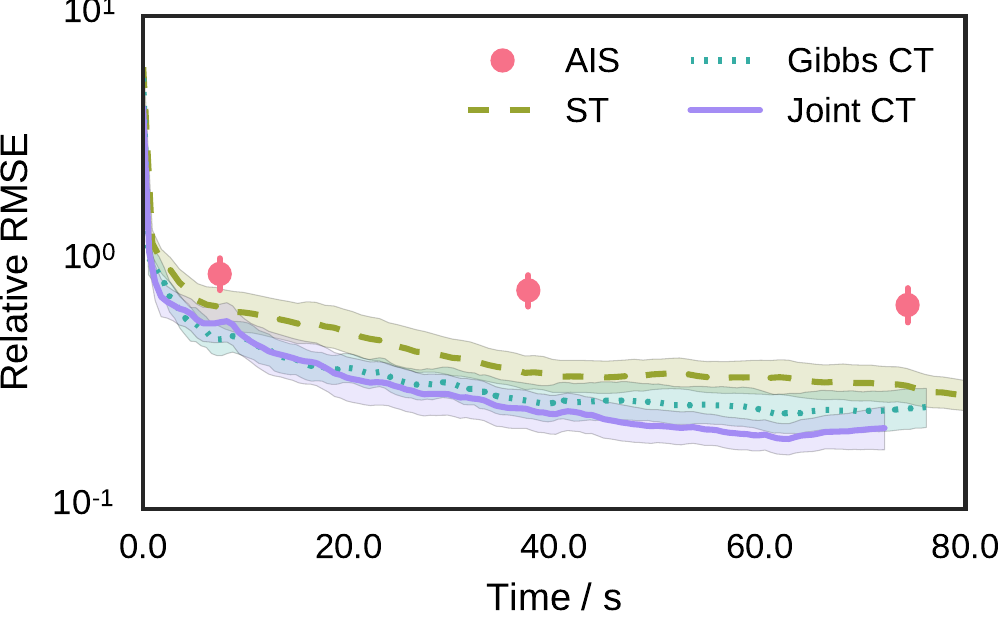}
\caption{$\expc{\rvct{x}}$}\label{sfig:bmr-30-unit-scale-6-mean}
\end{subfigure}
\begin{subfigure}[b]{.33\linewidth}
\vskip 0pt
\centering
\includegraphics[width=0.95\textwidth]{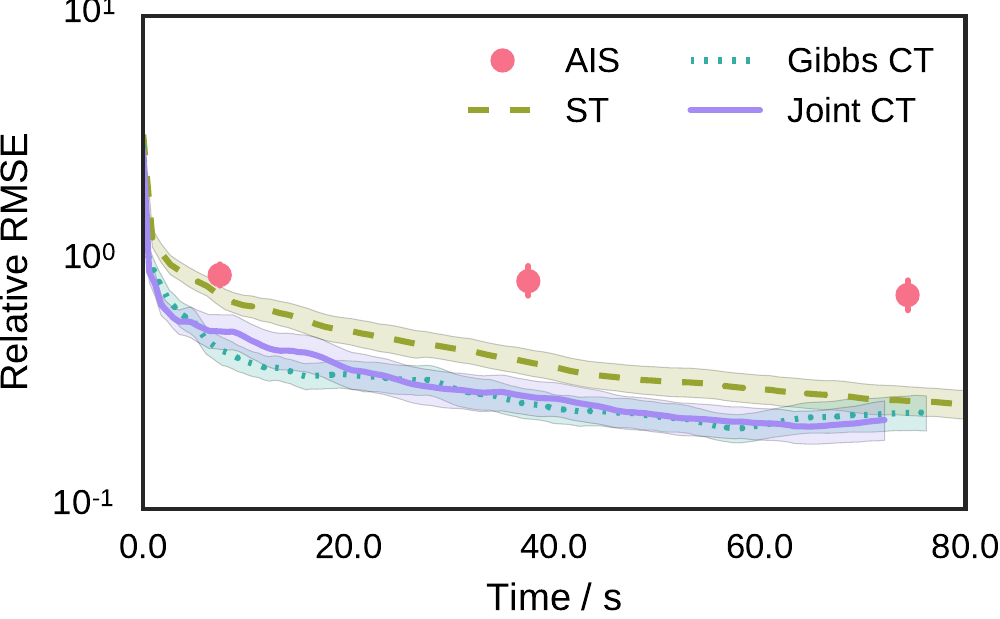} 
\caption{$\expc{\rvct{x}\rvct{x}\tr} - \expc{\rvct{x}}\expc{\rvct{x}}\tr$}\label{sfig:bmr-30-unit-scale-6-covar}
\end{subfigure}
\caption{\abbrdef{Root Mean Squared Errors}{RMSEs} in empirical moments estimated from \abbrref{MCMC} samples against run time for various thermodynamic ensemble \abbrref{MCMC} methods run on Gaussian Boltzmann machine relaxation target distributions. All RMSEs are relative to the RMSE of the corresponding approximate moments calculated using the moment-matched variational mixtures, so values below $1$ represent improvement on deterministic approximation. For \abbrref{AIS} points across time axis represent increasing number of inverse temperatures: $(1,\, 5,\,10)\times 10^3$. For \abbrref{ST}, Gibbs \abbrref{CT} and joint \abbrref{CT} curves show RMSEs for expectations calculated with increasing number of samples from chains. All curves / points show mean across 10 runs for each of 10 generated parameter sets. Filled regions / error bars show $\pm 3$ standard errors of mean.}
\label{fig:bmr-30-unit-scale-6-results}
\end{figure*}

For a second more challenging test case, we performed inference in Gaussian mixture relaxations of a set of ten synthetic Boltzmann machine distributions \citep{zhang2012continuous}. The parameters of the Boltzmann machine distributions were randomly generated so that the corresponding relaxations are highly multimodal and so challenging to explore well. 

The moments of the relaxation distributions can be calculated from the moments of the original discrete Boltzmann machine distribution, which for models with a small number of binary units $D_B$ (30 in our experiments) can be computed exactly by exhaustive iteration across the $2^{D_B}$ discrete states. This allows ground truth moments to be calculated against which convergence can be checked. The parametrisation used is described in in Appendix B. A Gaussian base density and approximate normalising constant $\zeta$ was fit to each the 10 relaxation target densities by matching moments to a mixture of variational Gaussian approximations (individually fitted using a mean-field approach based on the underlying Boltzmann machine distribution) as described in section \ref{sec:choosing-base-density}.

Plots showing the \abbrdef{root mean squared error}{RMSE} in estimates of $\log Z$ and the mean and covariance of the relaxation distribution against computational run time for different sampling methods are shown in Figure \ref{fig:bmr-30-unit-scale-6-results}. The \abbrref{RMSE} values are normalised by the \abbrref{RMSE}s of the corresponding estimated moments used in the base density (and $\log\zeta$) such that values below unity indicate an improvement in accuracy over the variational approximation. The curves shown are \abbrref{RMSE}s averaged over 10 independent runs for each of the 10 generated parameter sets, with the filled regions indicating $\pm 3$ standard errors of the mean. The free parameters of all methods were hand-tuned on one parameter set and these values then fixed across all runs. All methods used a shared Theano \citep{theano2016theano} implementation running on a Intel Core i5-2400 quad-core CPU for the underlying \abbrref{HMC} updates and so run times are roughly comparable.

For \abbrdef{simulated tempering}{ST}, \emph{Rao Blackwellised} estimators were used as described in \citep{carlson2016partition}, with \abbrref{HMC}-based updates of the target state $\rvct{x} \gvn \upbeta$ interleaved with independent sampling of $\upbeta \gvn \rvct{x}$, and 1000 $\beta_n$ values used. For \abbrref{AIS}, \abbrref{HMC} updates were used for the transition operators and separate runs with 1000, 5000 and 10000 $\beta_n$ values used to obtain estimates at different run times. For the tempering approaches run times correspond to increasing numbers of \abbrref{MCMC} samples.

The two \abbrdef{continuous tempering}{CT} approaches, Gibbs \abbrref{CT} and joint \abbrref{CT}, both dominate in terms of having lower average \abbrref{RMSEs} in all three moment estimates across all run times, with joint \abbrref{CT} showing marginally better performance on estimates of $\log\normconst$ and $\expc{\rvct{x}}$. The tempering approaches seem to outperform \abbrref{AIS} here, possibly as the highly multimodal nature of the target densities favours the ability of tempered dynamics to move up an down the inverse temperature scale and so in and out of modes in the target density, unlike \abbrref{AIS} where the deterministic temperature traversal is more likely to mean the runs end up being confined to a single mode after the initial transitions for low $\beta_n$.

\subsection{Bayesian hierarchical regression model}\label{subsec:exp-hier-regression}

\begin{figure*}[!ht]
\centering
\begin{subfigure}[b]{.5\linewidth}
\vskip 0pt
\centering
\scalebox{0.95}{
\tikz{ 
	\node[obs] (y) {$y^{(i)}$} ; %
	\factor[left=of y] {epsilon-y-yhat} {below:\tiny$\nrm{\hat{y}^{(i)},\,\epsilon^2}$} {} {}; %
	\node[latent, left=of epsilon-y-yhat, xshift=5mm, yshift=10mm] (epsilon) {$\epsilon$} ; %
	\node[latent, left=of epsilon-y-yhat, xshift=5mm] (yhat) {$\hat{y}^{(i)}$} ; %
	\factor[left=of epsilon] {pr-epsilon} {\tiny$\mathcal{C}_{\geq 0}(2.5)$} {} {}; %
	\factor[left=of yhat, xshift=-3mm] {alpha-yhat-beta}
	{below:\tiny$\delta\lpa\vct{\alpha}[c^{(i)}] + \vct{\beta}\tr\vct{x}^{(i)}\rpa$} {} {}; %
	\node[const, left=of alpha-yhat-beta, xshift=3mm, yshift=-6mm] (dummy) {} ; %
	\plate {platedata} {(y)(yhat)(alpha-yhat-beta)(dummy)} {\tiny$i\in\fset{1\dots N}$}; %
	\node[latent, above left=of alpha-yhat-beta, xshift=-5mm] (alpha) {$\vct{\alpha}$} ; %
	\node[latent, below left=of alpha-yhat-beta, xshift=-5mm] (beta) {$\vct{\beta}$} ; %
	\factor[left=of alpha, xshift=-2mm] {sigmaalpha-alpha-mualpha} {\tiny$\nrm{\mu_\alpha\vct{1},\,\sigma_\alpha^2\mtx{I}}$} {} {} ; %
	\node[latent, left=of sigmaalpha-alpha-mualpha, yshift=5mm] (sigmaalpha) {$\sigma_\alpha$} ; %
	\node[latent, left=of sigmaalpha-alpha-mualpha, yshift=-5mm] (mualpha) {$\mu_\alpha$} ; %
    \factor[left=of sigmaalpha] {pr-sigmaalpha} {\tiny$\mathcal{C}_{\geq 0}(2.5)$} {} {} ; %
    \factor[left=of mualpha] {pr-mualpha} {\tiny$\nrm{0,\,1}$} {} {} ; %
	\factor[left=of beta, xshift=-2mm] {sigmabeta-beta-mubeta} {\tiny$\nrm{\mu_\beta\vct{1},\,\sigma_\beta^2\mtx{I}}$} {} {} ; %
	\node[latent, left=of sigmabeta-beta-mubeta, yshift=5mm] (sigmabeta) {$\sigma_\beta$} ; %
	\node[latent, left=of sigmabeta-beta-mubeta, yshift=-5mm] (mubeta) {$\mu_\beta$} ; %
    \factor[left=of sigmabeta] {pr-sigmabeta} {\tiny$\mathcal{C}_{\geq 0}(2.5)$} {} {} ; %
    \factor[left=of mubeta] {pr-mubeta} {\tiny$\nrm{0,\,1}$} {} {} ; %
	\edge {epsilon} {epsilon-y-yhat} ; %
	\edge {yhat} {epsilon-y-yhat} ; %
	\edge {epsilon-y-yhat} {y} ; %
	\edge {pr-epsilon} {epsilon} ; %
	\edge {alpha} {alpha-yhat-beta} ; %
	\draw[->] (beta) to[bend left=30] (alpha-yhat-beta);
	\edge {alpha-yhat-beta} {yhat} ; %
	\edge {sigmaalpha} {sigmaalpha-alpha-mualpha} ; %
	\edge {mualpha} {sigmaalpha-alpha-mualpha} ; %
	\edge {sigmaalpha-alpha-mualpha} {alpha} ; %
	\edge {pr-sigmaalpha} {sigmaalpha} ; %
	\edge {pr-mualpha} {mualpha} ; %
	\edge {sigmabeta} {sigmabeta-beta-mubeta} ; %
	\edge {mubeta} {sigmabeta-beta-mubeta} ; %
	\edge {sigmabeta-beta-mubeta} {beta} ; %
	\edge {pr-sigmabeta} {sigmabeta} ; %
	\edge {pr-mubeta} {mubeta} ; %
}
}
\vskip 5pt
\caption{Factor graph of hierarchical model.}
\label{sfig:hier-lin-regression-factor}
\end{subfigure}%
\begin{subfigure}[b]{.5\linewidth}
\vskip 0pt
\centering
\includegraphics[width=0.95\textwidth]{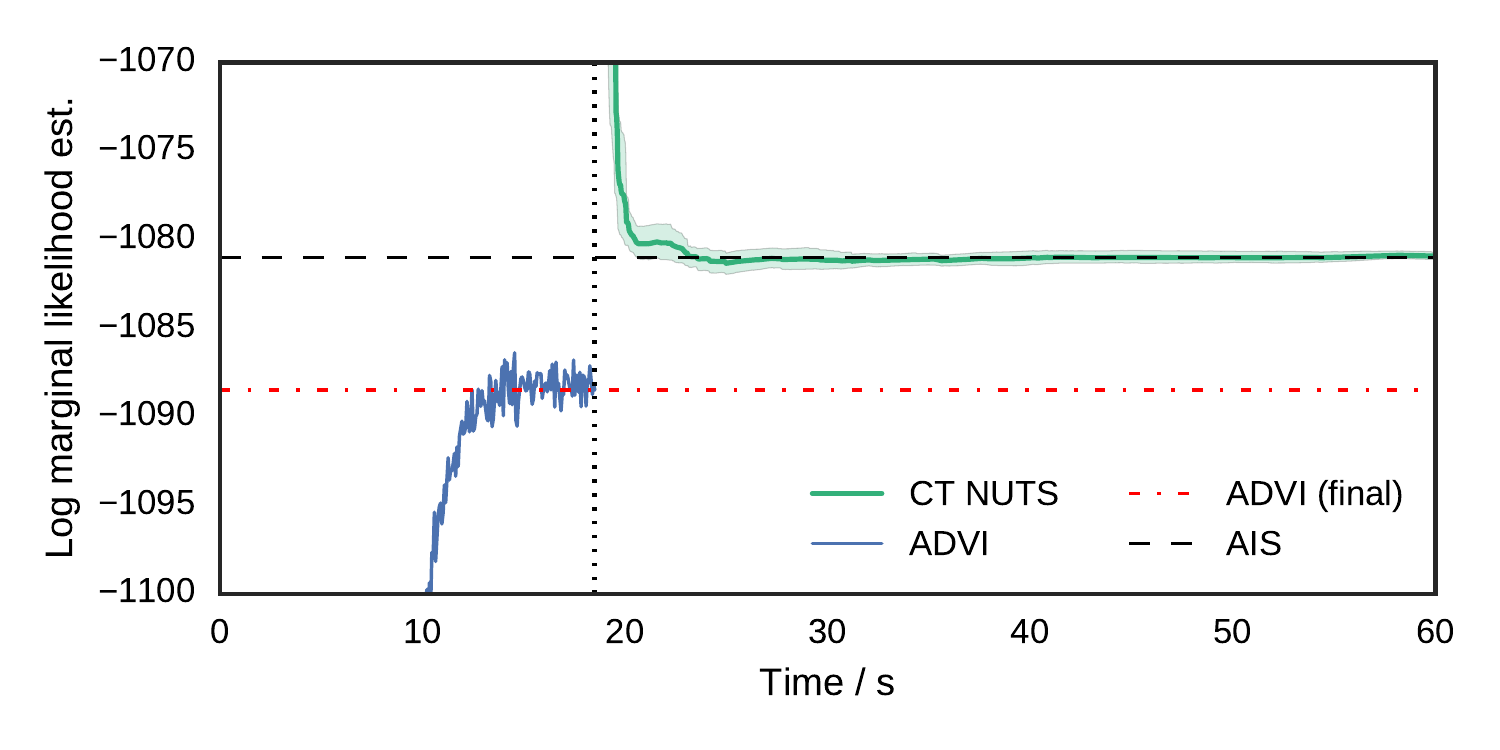}
\caption{Log marginal likelihood estimates.}\label{sfig:hier-lin-regression-marg-lik}
\vskip 0pt
\end{subfigure}
\caption{\subref{sfig:hier-lin-regression-factor} Factor graph for hierarchical regression model for \emph{Radon} data. Factor notation - $\nrm{\mu,\,\sigma}$: normal distribution with mean $\mu$, (co-)variance $\sigma$; $\mathcal{C}_{\geq 0}(\gamma)$ half-Cauchy distribution with scale $\gamma$; $\delta(z)$ Dirac-delta located at $z$. \subref{sfig:hier-lin-regression-marg-lik} Log marginal likelihood estimates against run time for hierarchical regression model. Black dashed line shows estimated log marginal likelihood from a long \abbrref{AIS} run which is used as a proxy for the ground truth. The noisy blue curve shows the \emph{evidence lower bound} \abbrref{ADVI} variational objective over training and the red dot-dashed line the final converged value used for $\log\zeta$. The green curve shows log marginal likelihood estimates using samples from \abbrref{NUTS} chains running on the extended joint density in the estimator \eqref{eq:ct-norm-const-estimator}, with the run time corresponding to increasing samples being included in the estimator (offset by initial ADVI run time). Curve shows mean over 10 independent runs and filled region $\pm 3$ standard errors of mean.}
\label{fig:hier-lin-regression}
\end{figure*}

As a third experiment, we apply our joint continuous tempering approach to perform Bayesian inference in a hierarchical regression model for predicting indoor radon measurements \citep{gelman2006data}. To illustrate the ease of integrating our approach in existing \abbrref{HMC}-based inference software, this experiment was performed with the Python package PyMC3 \citep{salvatier2016probabilistic}, with its \abbrref{ADVI} feature used to fit the base density and its implementation of the adaptive
\abbrdef{No U-Turn Sampler}{NUTS} \citep{hoffman2014no} \abbrref{HMC} variant used to sample from the extended space.

The regression target in the dataset is measurements of the amount of radon gas $\rvar{y}^{(i)}$ in $N=919$ households. Two continuous regressors $\vct{x}^{(i)}$ and one categorical regressor $c^{(i)}$ are provided per household. A multilevel regression model defined by the factor graph in \ref{sfig:hier-lin-regression-factor} was used. The model includes five scalar parameters $(\upsigma_\alpha,\,\upmu_\alpha,\,\upsigma_\beta,\,\upmu_\beta,\,\upepsilon)$, an 85-dimensional intercept vector $\vct{\upalpha}$ and a two-dimensional regressor coefficients vector $\vct{\upbeta}$, giving 92 parameters in total.

As an example task, we consider inferring the marginal likelihood of the data under the model. Estimation of the marginal likelihood from \abbrref{MCMC} samples of the target density alone is non-trivial, with approaches such as the harmonic mean estimator having high variance. Here we try to establish if our approach can be used in a black-box fashion to compute a reasonable estimate of the marginal likelihood.

As our `ground truth' we use a large batch of long \abbrref{AIS} runs (average across 100 runs of 10000 inverse temperatures) on a separate Theano implementation of the model. We use \abbrref{ADVI} to fit a diagonal covariance Gaussian variational approximation to the target density and use this as the base density. \abbrref{NUTS} chains, initialised at samples from the base density, were then run on the extended space for 2500 iterations. The samples from these chain were used to compute estimates of the normalising constant (marginal likelihood) using the estimator \eqref{eq:ct-norm-const-estimator}. The results are shown in Figure \ref{fig:hier-lin-regression}. It can be seen that estimates from the \abbrref{NUTS} chains in the extended continuously tempered space quickly converge to a marginal likelihood estimate very close to the \abbrref{AIS} estimate, and significantly improve over the final lower bound on the marginal likelihood that \abbrref{ADVI} converges to.

\subsection{Importance weighted autoencoder image models}\label{subsec:exp-iwae}

\begin{figure*}[!ht]
\centering
\begin{subfigure}[t]{.5\linewidth}
\vskip 0pt
\centering
\includegraphics[width=0.95\textwidth]{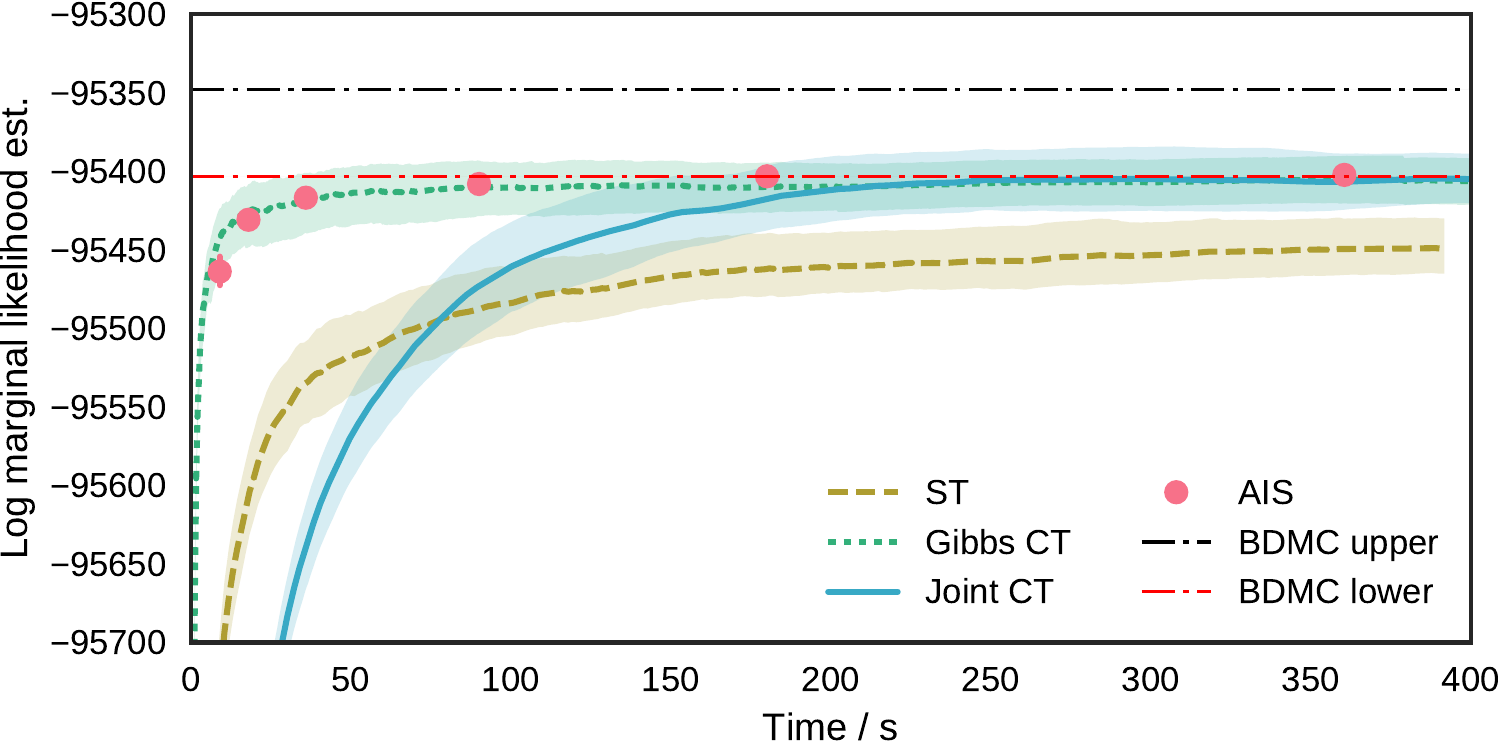}
\caption{MNIST log marginal likelihood estimates.}\label{sfig:mnist-log-marg-lik}
\end{subfigure}%
\begin{subfigure}[t]{.5\linewidth}
\vskip 0pt
\centering
\includegraphics[width=0.95\textwidth]{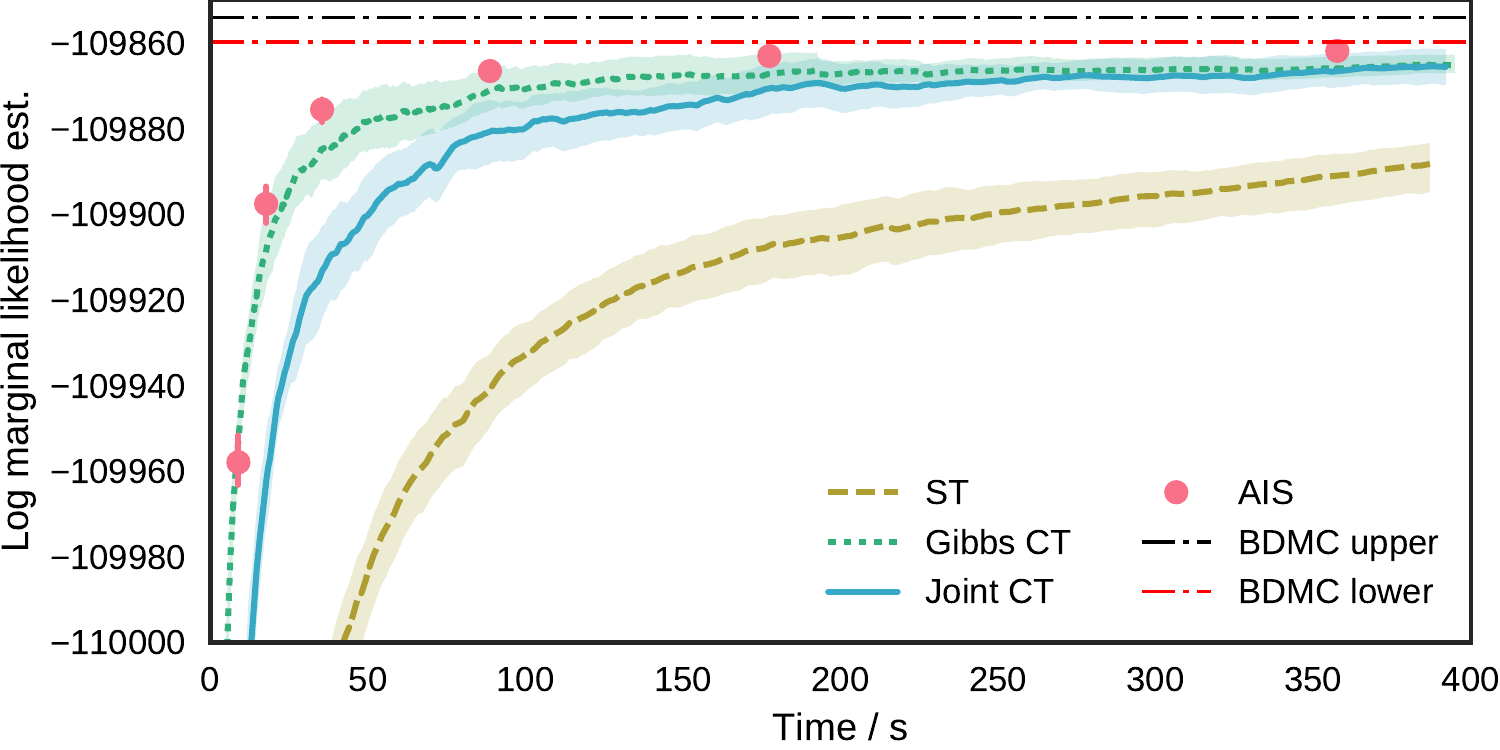}
\caption{Omniglot log marginal likelihood estimates.}\label{sfig:omni-log-marg-lik}
\vskip 0pt
\end{subfigure}
\caption{Estimates of the log joint marginal likelihood of 1000 generated images under the Bernoulli decoder distributions of two \abbrref{IWAE} models trained on the MNIST and Omniglot datasets against computation time. The black / red dashed lines show stochastic upper / lower bounds calculated using long \abbrref{BDMC} runs. For \abbrref{AIS} points across time axis represent increasing number of inverse temperatures: $(50,\,100,\,200,\,500,\,1000,\,2000)$. For \abbrref{ST}, Gibbs \abbrref{CT} and joint \abbrref{CT} curves show estimates calculated with increasing number of samples from chains. All curves / points show mean across 10 runs. 
}
\label{fig:iwae-marginal-likelihood-results}
\end{figure*}

\vspace{-1pt}
For our final experiments, we compare the efficiency of our continuous tempering approaches to simulated tempering and \abbrref{AIS} for marginal likelihood estimation in decoder-based generative models for images. Use of \abbrref{AIS} in this context was recently proposed in \citep{wu2016quantitative}. 

Specifically we estimate the joint marginal likelihood of 1000 generated binary images under the Bernoulli decoder distribution of two \abbrdef{importance weighted autoencoder}{IWAE} \citep{burda2016importance} models. Each \abbrref{IWAE} model has one stochastic hidden layer and a 50-dimensional latent space, with the two models trained on binarised versions of the MNIST \citep{lecun1998gradient} and Omniglot \citep{lake2015human} datasets using the code at \url{https://github.com/yburda/iwae}. The generated images used in the experiments are shown in Appendix C.

By performing inference on the per-image posterior densities on the latent representation given image, the joint marginal likelihood of the images can be estimated as the product of estimates of the normalising constants of the individual posterior densities. The use of generated images allows \abbrdef{bidirectional Monte Carlo}{BDMC} \citep{grosse2015sandwiching} to be used to `sandwich' the marginal likelihood with both stochastic upper and lower bounds formed with long forward and backward \abbrref{AIS} runs (averages over 16 independent runs with 10000 inverse temperatures as used in \citep{wu2016quantitative}). 

As the per-image latent representations are conditionally independent given the images, chains on all the posterior densities can be run in parallel, with the experiments in this section run on a NVIDIA Tesla K40 GPU to exploit this inherent parallelism. The encoder of the trained \abbrref{IWAE} models is an inference network which outputs the mean and diagonal covariance of a Gaussian variational approximation to the posterior density on the latent representation given an image and so was used to define per-image Gaussian base densities as suggested in \citep{wu2016quantitative}. Similarly the per-image $\log \zeta$ values were set using importance-weighted variational lower bound estimates for the per-image marginal likelihoods.

The results are shown in Figure \ref{fig:iwae-marginal-likelihood-results}, with the curves / points showing average results across 10 independent runs and filled regions / bars $\pm 3$ standard error of means for the estimates. Here Gibbs \abbrref{CT} and \abbrref{AIS} perform similarly, with joint \abbrref{CT} converging less quickly and simulated tempering significantly less efficient. The quick convergence of \abbrref{AIS} and Gibbs \abbrref{CT} here suggests the posterior densities are relatively easy for the dynamics to explore and well matched by the Gaussian base densities, limiting the gains from any more coherent exploration of the extended space by the joint \abbrref{CT} updates. The higher per-leapfrog-step costs of the \abbrref{HMC} updates in the extended space therefore mean the joint \abbrref{CT} approach is less efficient overall here. The poorer performance of simulated tempering here is in part due to the generation of the discrete random indices becoming a bottleneck in the \abbrref{GPU} implementation.

A possible reason for the better relative performance of \abbrref{AIS} here compared to the experiments in Section \ref{subsec:exp-bm-relaxations} is its more effective utilisation of the parallel compute cores available when running for example on a \abbrref{GPU}. Multiple \abbrref{AIS} chains can be run for each data point and then the resulting unbiased estimates for each data points marginal likelihood averaged (reducing the per data point variance) before taking their product for the joint marginal likelihood estimate. While it is also possible to run multiple tempered chains per data point and similarly combine the estimates, empirically we found that greater gains in estimation accuracy came from running a single longer chain rather than multiple shorter chains of total length equivalent to the longer chain. This can be explained by the initial \emph{warm up} transients of each shorter chain having a greater biasing effect on the overall estimate compared to running longer chains. 

\section{Discussion}

The approach we have presented is a simple but powerful extension to existing tempering methods which can both help exploration of distributions with multiple isolated modes and allow estimation of the normalisation constant of the target distribution. A key advantage of the joint continuous tempering method is the simplicity of its implementation - it simply requires running \abbrref{HMC} in an extended state space and so can easily be used for example within existing probabilistic programming software which use \abbrref{HMC}-based methods for inference such as PyMC3 \citep{salvatier2016probabilistic} and Stan \citep{carpenter2016stan}. By updating the inverse temperature jointly with the original target state, it is also possible to leverage adaptive \abbrref{HMC} variants such as \abbrref{NUTS} \citep{hoffman2014no} to perform tempered inference in a `black-box' manner without the need to separately tune the updates of the inverse temperature variable.

The Gibbs continuous tempering method also provides a relatively black-box framework for tempering. Compared to simulated tempering it removes the requirement to choose the number and spacing of discrete inverse temperatures and also replaces generation of a discrete random variate from a categorical distribution when updating $\upbeta$ given $\rvct{x}$ (which as seen in Section \ref{subsec:exp-iwae} can become a computational bottleneck) with generation of a truncated exponential variate (which can be performed efficiently by inverse transform sampling). Compared to the joint continuous tempering approach, the Gibbs approach is less simple to directly integrate in to existing \abbrref{HMC} implementations due to the separate $\upbeta$ updates, but eliminates the need to tune the temperature control mass value $m$ and achieved similar or better sampling efficiency in the experiments in Section \ref{sec:experiments}.

Our proposal to use variational approximations within an \abbrref{MCMC} framework can be viewed within the context of several existing approaches which suggest combining variational and \abbrref{MCMC} inference methods. \emph{Variational MCMC} \citep{de2001variational} proposes using a variational approximation as the basis for a proposal distribution in a Metropolis-Hastings \abbrref{MCMC} method. \emph{MCMC and Variational Inference: Bridging the Gap} \citep{salimans2015markov} includes parametrised \abbrref{MCMC} transitions within a (stochastic) variational approximation and optimises the variational bound over these (and a base distribution's) parameters. Here we exploit cheap (relative to running a long \abbrref{MCMC} chain) but biased variational approximations to a target distribution and its normalising constant, and propose using them within an \abbrref{MCMC} method which gives asymptotically exact results to help improve sampling efficiency.

\subsubsection*{Acknowledgements}

We thank Ben Leimkuhler for introducing us to the extended Hamiltonian continuous tempering approach and to the anonymous reviewers of an earlier workshop version of this paper for their useful feedback. This work was supported in part by grants EP/F500386/1 and BB/F529254/1 for the University of Edinburgh School of Informatics Doctoral Training Centre in Neuroinformatics and Computational Neuroscience (www.anc.ac.uk/dtc) from the UK Engineering and Physical Sciences Research Council (EPSRC), UK Biotechnology and Biological Sciences Research Council (BBSRC), and the UK Medical Research Council (MRC).  

{
\bibliography{refs}
}

\newpage
\onecolumn
\appendix

\section{Bounding the inverse temperature marginal density}\label{app:bounding-the-partition-function}

We have a joint density on $(\rvct{x},\,\upbeta)$
\begin{equation}
\pden{\rvct{x}=\vct{x},\,\upbeta = \beta} =
\frac{1}{C} \exp\lsb -\beta\phi(\vct{x}) -\beta\log\zeta -(1-\beta)\psi(\vct{x})\rsb.
\label{eq:x-beta-joint-density-app}
\end{equation}
The resulting marginal density on $\upbeta$ is
\begin{equation}\label{eq:beta-marginal-density}
\pden{\upbeta=\beta} = \int_{\set{X}} \pden{\rvct{x}=\vct{x},\,\upbeta = \beta} \,\dr\vct{x}
= \frac{1}{C\zeta^\beta} \int_{\set{X}} \exp\lsb -\beta\phi(\vct{x}) -(1-\beta)\psi(\vct{x})\rsb\,\dr\vct{x}.
\end{equation}
To derive an upper-bound on $\pden{\upbeta=\beta}$ we use H\"older's inequality
\begin{equation}\label{eq:holders-inequality}
  \int_{\set{X}} g(\vct{x}) h(\vct{x}) \,\dr\vct{x} \leq 
  \lpa \int_{\set{X}} |g(\vct{x})|^{\frac{1}{a}} \,\dr\vct{x} \rpa^{a}
  \lpa \int_{\set{X}} |h(\vct{x})|^{\frac{1}{1-a}} \,\dr\vct{x} \rpa^{1-a}
\end{equation}
where $a \in [0,\,1]$ and $g$ and $h$ are measurable functions. We also use the definitions 
\begin{equation}\label{eq:target-and-based-norm-consts}
  \int_\set{X} \exp\lsb-\phi(\vct{x})\rsb\,\dr\vct{x} = Z
  \quad\text{and}\quad
  \int_\set{X} \exp\lsb-\psi(\vct{x})\rsb\,\dr\vct{x} = 1.
\end{equation}
From \eqref{eq:beta-marginal-density} we have that
\begin{equation}
  \pden{\upbeta=\beta}
  =
  \frac{1}{C\zeta^{\beta}} 
  \int_\set{X} 
    \lpa \exp\lsb-\phi(\vct{x}) \rsb^{\beta}\rpa \lpa \exp\lsb-\psi(\vct{x})\rsb^{1 - \beta}\rpa
  \,\dr\vct{x}.
\end{equation}
Applying H\"older's inequality \eqref{eq:holders-inequality} with $g(\vct{x}) = \exp[-\phi(\vct{x})]^{\beta}$, $h(\vct{x}) = \exp[-\psi(\vct{x})]^{1-\beta}$ and $a = \beta$
\begin{align}
  \pden{\upbeta=\beta}
  &\leq
  \frac{1}{C\zeta^{\beta}} 
  \lpa
  \int_\set{X} 
    \left| \exp\lsb-\phi(\vct{x}) \rsb^{\beta} \right|^{\frac{1}{\beta}}
  \,\dr\vct{x}
  \rpa^{\beta}
  \lpa
  \int_\set{X} 
    \left|\exp\lsb-\psi(\vct{x})\rsb^{1 - \beta}\right|^{\frac{1}{1-\beta}}
  \,\dr\vct{x}
  \rpa^{1-\beta}  
  \\
  &=
  \frac{1}{C\zeta^{\beta}} 
  \lpa
  \int_\set{X} 
    \exp\lsb-\phi(\vct{x})\rsb
  \,\dr\vct{x}
  \rpa^{\beta}
  \lpa
  \int_\set{X} 
    \exp\lsb-\psi(\vct{x})\rsb
  \,\dr\vct{x}
  \rpa^{1-\beta}.
\end{align}
Substituting the definitions in \eqref{eq:target-and-based-norm-consts} gives
\begin{equation}
  \pden{\upbeta = \beta} \leq \frac{1}{C} \lpa \frac{Z}{\zeta} \rpa^\beta.
\end{equation}
To derive a lower-bound on $\pden{\upbeta=\beta}$, we use Jensen's inequality
\begin{equation}\label{eq:jensens-inequality}
  \varphi\lpa \int_{\set{X}} g(\vct{x}) q(\vct{x}) \,\dr\vct{x}\rpa \geq
 \int_{\set{X}} \varphi\lpa g(\vct{x}) \rpa q(\vct{x}) \,\dr\vct{x},
\end{equation}
for a concave function $\varphi$, normalised density $q : \int_\set{X} q(\vct{x}) \,\dr\vct{x} = 1$ and measurable $g$. The logarithm of \eqref{eq:beta-marginal-density} gives
\begin{equation}
  \log\pden{\upbeta = \beta} + \beta\log\zeta + \log C
  =
  \log\lpa
  \int_\set{X} 
    \exp\lpa-\beta\lsb\phi(\vct{x}) - \psi(\vct{x})\rsb\rpa \exp\lsb-\psi(\vct{x})\rsb
  \,\dr\vct{x}
  \rpa.
\end{equation}
Applying Jensen's inequality \eqref{eq:jensens-inequality} with $\varphi = \log$, $q = \exp(-\psi)$ and $g = \exp\lsb-\beta\lpa\phi - \psi\rpa\rsb$
\begin{align}
  \log\pden{\upbeta = \beta} + \log C + \beta\log\zeta
  &\geq
  \beta
  \int_\set{X} 
    \lsb\psi(\vct{x}) - \phi(\vct{x})\rsb \exp\lsb-\psi(\vct{x})\rsb
  \,\dr\vct{x}\\
  &=
  \beta
  \int_\set{X} 
    \lpa
    \log Z - \log Z -
    \log\exp\lsb-\psi(\vct{x}) + \phi(\vct{x})\rsb
    \rpa 
    \exp\lsb-\psi(\vct{x})\rsb
  \,\dr\vct{x}
  \\ 
  &=
  \beta\log Z
  -\beta
  \int_\set{X} 
    \exp\lsb-\psi(\vct{x})\rsb \log\lpa\frac{\exp\lsb-\psi(\vct{x})\rsb}{\exp\lsb-\phi(\vct{x})\rsb / Z}\rpa
  \,\dr\vct{x}.
\end{align}
Recognising the integral in the last line as the \abbrdef{Kullback--Leibler}{KL} divergence $d^{b\to t}$ from the base density $\exp\lsb-\psi(\vct{x})\rsb$ to the target density $\exp\lsb-\phi(\vct{x})\rsb / Z$
\begin{equation}
  d^{b\to t} = \int_{\set{X}} 
    \exp\lsb -\psi(\vct{x})\rsb \log\lpa\frac{\exp\lsb -\psi(\vct{x})\rsb}{\exp\lsb -\phi(\vct{x})\rsb / Z}\rpa
  \,\dr\vct{x},
\end{equation} 
and taking the exponential of both sides and rearranging we have
\begin{equation}
  \pden{\upbeta = \beta} \geq
  \frac{1}{C} \lpa \frac{Z}{\zeta} \rpa^\beta \exp\lpa -\beta d^{b\to t}\rpa.
\end{equation}
By instead noting \eqref{eq:beta-marginal-density} can be rearranged into the form
\begin{equation}
  \log\pden{\upbeta = \beta} + \log C + \beta\log\zeta - \log Z
  =
  \log\lpa
  \int_\set{X} 
    \exp\lpa-(1-\beta)\lsb\psi(\vct{x}) - \phi(\vct{x})\rsb\rpa \frac{1}{Z}\exp\lsb-\phi(\vct{x})\rsb
  \,\dr\vct{x}
  \rpa,
\end{equation}
by an equivalent series of steps we can also derive a bound using the reversed form of the \abbrref{KL} divergence 
\begin{equation}
  d^{t\to b} = \int_{\set{X}} 
    \frac{1}{Z}\exp\lsb -\phi(\vct{x})\rsb \log\lpa\frac{\exp\lsb -\phi(\vct{x})\rsb / Z}{\exp\lsb -\psi(\vct{x})\rsb}\rpa
  \,\dr\vct{x}.
\end{equation} 
from the target to the base distribution, giving that
\begin{equation}
  \pden{\upbeta = \beta} \geq 
  \frac{1}{C} \lpa \frac{Z}{\zeta} \rpa^\beta \exp\lsb -(1 - \beta) d^{t\to b}\rsb.
\end{equation}

\section{Gaussian mixture Boltzmann machine relaxation}\label{app:boltzmann-machine-relaxation}

We define a \emph{Boltzmann machine distribution} on a signed binary state $\rvct{s} \in \fset{-1,\,+1}^{D_B} = \set{S}$ as
\begin{equation}
  \prob{\rvct{s} = \vct{s}} = 
  \frac{1}{Z_B} \exp\lpa\frac{1}{2}\vct{s}\tr\mtx{W}\vct{s} + \vct{s}\tr\vct{b}\rpa
  \qquad
  Z_B = \sum_{\vct{s}\in\set{S}} \lsb\exp\lpa\frac{1}{2}\vct{s}\tr\mtx{W}\vct{s} + \vct{s}\tr\vct{b}\rpa\rsb.
\end{equation}
\noindent
We introduce an auxiliary real-valued vector random variable $\rvct{x}\in\reals^D$ with a Gaussian conditional distribution
\begin{equation}
  \pden{\rvct{x} = \vct{x} \gvn \rvct{s} = \vct{s}} =
  \frac{1}{(2\pi)^{\nicefrac{D}{2}}} \exp\lsb 
    -\frac{1}{2} \lpa \vct{x} - \mtx{Q}\tr\vct{s} \rpa \tr \lpa \vct{x} - \mtx{Q}\tr\vct{s} \rpa 
  \rsb
\end{equation}
with $\mtx{Q}$ a $D_B \times D$ matrix such that $\mtx{Q}\mtx{Q}\tr = \mtx{W} + \mtx{D}$ for some diagonal $\mtx{D}$ which makes $\mtx{W} + \mtx{D}$ positive semi-definite. In our experiments, based on the observation in \citep{zhang2012continuous} that minimising the maximum eigenvalue of $\mtx{W} + \mtx{D}$ decreases the maximal separation between the Gaussian components in the relaxation, we set $\mtx{D}$ as the solution to the semi-definite programme 
\begin{equation}
  \min_{\mtx{D}} \lsb \lambda_{\textsc{max}}\lpa \mtx{W} + \mtx{D} \rpa \rsb 
  : \mtx{W} + \mtx{D} \succeq 0
\end{equation}
where $\lambda_{\textsc{max}}$ denotes the maximal eigenvalue. In general the optimised $\mtx{W} + \mtx{D}$ lies on the semi-definite cone and so has rank less than $D_B$ hence a $\mtx{Q}$ can be found such that $D < D_B$.
\noindent
The resulting joint distribution on $(\rvct{x},\,\rvct{s})$ is
\begin{align}
  \pden{\rvct{x} = \vct{x},\, \rvct{s} = \vct{s}} 
  &=
  \frac{1}{(2\pi)^{\nicefrac{D}{2}} Z_B} \exp\lsb 
    -\frac{1}{2} \vct{x}\tr\vct{x} + \vct{s}\tr\mtx{Q}\vct{x} 
    -\frac{1}{2}\vct{s}\tr\mtx{Q}\mtx{Q}\tr\vct{s} + \frac{1}{2}\vct{s}\tr\mtx{W}\vct{s} 
    + \vct{s}\tr\vct{b}
  \rsb
  \\
  &=
  \frac{1}{(2\pi)^{\nicefrac{D}{2}} Z_B} \exp\lsb 
    -\frac{1}{2} \vct{x}\tr\vct{x} + \vct{s}\tr\lpa \mtx{Q}\vct{x} + \vct{b} \rpa
    -\frac{1}{2}\vct{s}\tr\mtx{D}\vct{s}
  \rsb
  \\
  &=
  \frac{1}{(2\pi)^{\nicefrac{D}{2}} Z_B \exp\lpa\frac{1}{2}\Tr(\mtx{D})\rpa} 
  \exp\lsb -\frac{1}{2} \vct{x}\tr\vct{x} \rsb
  \prod_{i=1}^{D_B} \lpa
  \exp\lsb
    s_i\lpa \vct{q}_i\tr\vct{x} + b_i \rpa
  \rsb
  \rpa,
\end{align}
where $\fset{\vct{q}_i\tr}_{i=1}^{D_B}$ are the $D_B$ rows of $\mtx{Q}$.
\noindent
We can marginalise over the binary state $\rvct{s}$ as each $\rvar{s}_i$ is conditionally independent of all the others given $\rvct{x}$ in the joint distribution. This gives the \emph{Boltzmann machine relaxation} density on $\rvct{x}$
\begin{equation}
  \pden{\rvct{x} = \vct{x}} =
  \frac{2^{D_B}}{(2\pi)^{\nicefrac{D}{2}} Z_B \exp\lpa\frac{1}{2}\Tr(\mtx{D})\rpa} 
  \exp\lpa -\frac{1}{2} \vct{x}\tr\vct{x} \rpa 
  \prod_{i=1}^{D_B} \lsb \cosh\lpa\vct{q}_i\tr\vct{x} + b_i \rpa\rsb,
\end{equation}
which is a specially structured Gaussian mixture density with $2^{D_B}$ components.
\noindent
If we define $\pden{\rvct{x}=\vct{x}} = \frac{1}{Z} \exp\lsb-\phi(\vct{x})\rsb$ with
\begin{equation}
  \phi(\vct{x}) = 
  \frac{1}{2} \vct{x}\tr\vct{x} -
  \sum_{i=1}^{D_B} \lsb \log\cosh\lpa\vct{q}_i\tr\vct{x} + b_i \rpa\rsb,
\end{equation}
then the normalisation constant $Z$ of the relaxation density can be related to the normalising constant of the corresponding Boltzmann machine distribution by
\begin{equation}
  \log{Z} = \log Z_B + \frac{1}{2}\Tr(\mtx{D}) + \frac{D}{2}\log(2\pi) - D_B\log 2.
\end{equation}
\noindent
It can also be shown that the first and second moments of the relaxation distribution are related to the first and second moments of the corresponding Boltzmann machine distribution by
\begin{align}
  \expc{\rvct{x}} 
  = \int_{\set{X}} \vct{x} \sum_{\vct{s}\in\set{S}} \lpa 
    \pden{\rvct{x} = \vct{x} \gvn \rvct{s} = \vct{s}} \prob{\rvct{s} = \vct{s}} 
  \rpa 
  \,\dr\vct{x} 
  = \sum_{\vct{s}\in\set{S}} \lpa 
    \int_{\set{X}} \vct{x} \,\nrm{\vct{x}; \mtx{Q}\tr\vct{s},\,\mtx{I}} \,\dr\vct{x}
    \prob{\rvct{s} = \vct{s}} 
  \rpa
  = \expc{
    \mtx{Q}\tr\rvct{s}
  }
  =
  \mtx{Q}\tr \expc{\rvct{s}},
  \\
  \textrm{and }
  \expc{\rvct{x}\rvct{x}\tr} 
  = \sum_{\vct{s}\in\set{S}} \lpa 
    \int_{\set{X}} \vct{x}\vct{x}\tr \,\nrm{\vct{x}; \mtx{Q}\tr\vct{s},\,\mtx{I}} \,\dr\vct{x}
    \prob{\rvct{s} = \vct{s}} 
  \rpa
  = \expc{\mtx{Q}\tr\rvct{s}\rvct{s}\mtx{Q} + \mtx{I}}
  = \mtx{Q}\tr\expc{\rvct{s}\rvct{s}\tr}\mtx{Q} + \mtx{I}.
\end{align}

\begin{figure}[!ht]
\centering
\includegraphics[width=0.9\textwidth]{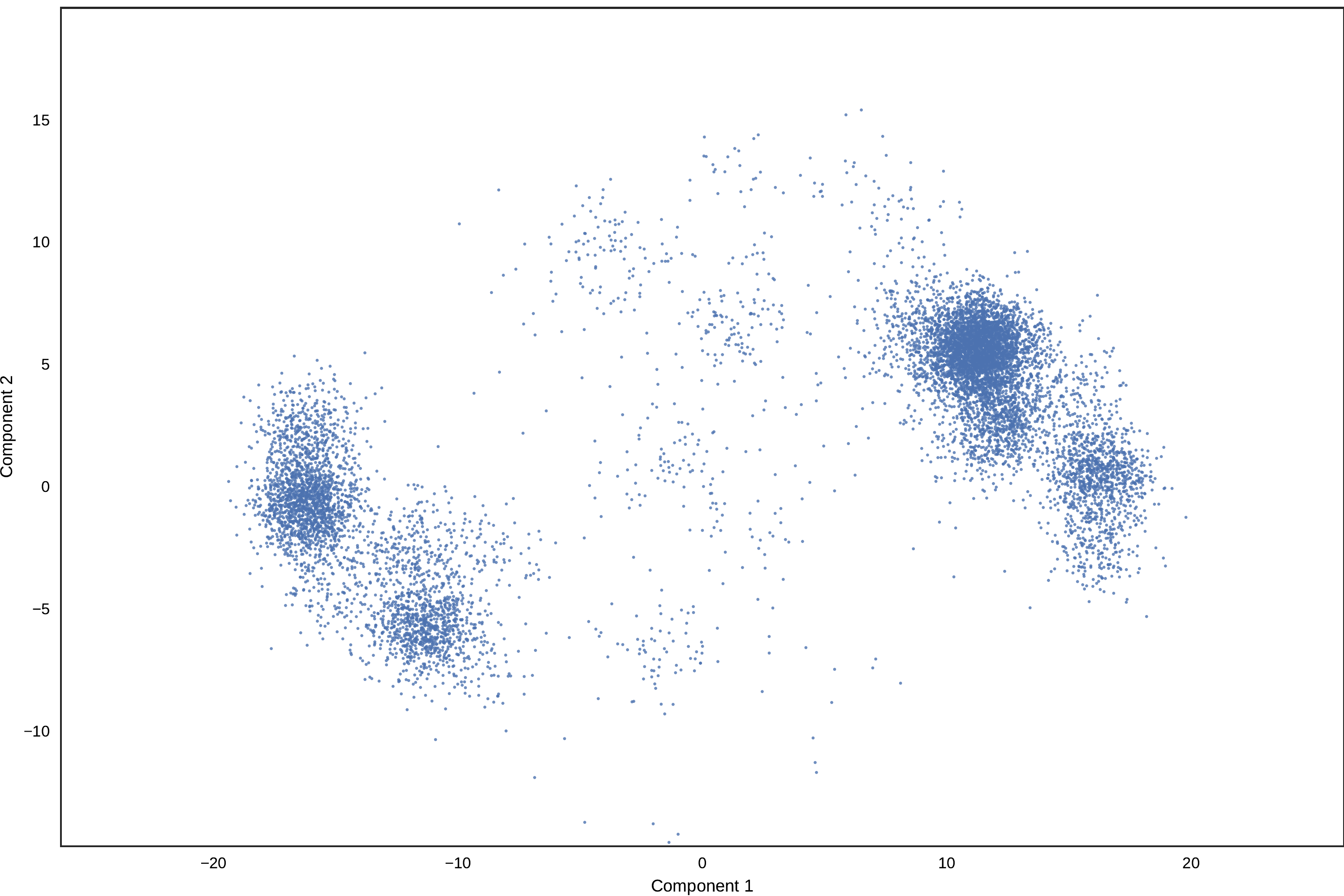}
\caption{Two-dimensional projection of 10000 independent samples from a Gaussian mixture relaxation of a Boltzmann machine distribution. The parameters $\mtx{W}$ and $\vct{b}$ of the Boltzmann machine distribution where generated as described in Section \ref{app:boltzmann-machine-relaxation}, with here $D_B = 28$ (rather than $D_B = 30$ as in the experiments) as independent sampling from larger systems exceeded the memory available on the workstation used. The two components shown correspond to the two eigenvectors of the generated basis $\mtx{R}$ with the largest corresponding eigenvalues.}
\label{fig:bmr-samples-2d}
\end{figure}

The weight parameters $\mtx{W}$ of the Boltzmann machine distributions used in the experiments in Section \ref{subsec:exp-bm-relaxations} were generated using an eigendecomposition based method. A uniformly distributed (with respect to the Haar measure) random orthogonal matrix $\mtx{R}$ was sampled. A vector of eigenvalues $\vct{e}$ was generated by sampling independent zero-mean unit-variance normal variates $n_i \sim \nrm{\cdot;\,0,\,1} ~\forall i \in \fset{1,\dots D_B}$ and then setting $e_i = s_1 \tanh(s_2 n_i) ~\forall i \in \fset{1,\dots D_B}$, with $s_1 = 6$ and $s_2 = 2$ in the experiments. This generates eigenvalues concentrated near $\pm s_1$ with this empirically observed to lead to systems which tended to be highly multimodal. A symmetric matrix $\mtx{V} = \mtx{R}\diag(\vct{e})\mtx{R}\tr$ was then computed and the weights $\mtx{W}$ set such that $W_{i,j} = V_{i,j} ~\forall i \neq j$ and $W_{i,i} = 0 ~\forall i \in \fset{1,\dots D_B}$. The biases $\vct{b}$ where generated using $b_i \sim \nrm{\cdot;\,0,0.1^2} ~\forall i \in \fset{1,\dots D_B}$. An example of a two-dimensional projection of independent samples from a Boltzmann machine relaxation density with $D=27$ ($D_B=28$), and $\mtx{W}$ and $\vct{b}$ generated as just described in shown in figure \ref{fig:bmr-samples-2d}. As can be seen even when projected down to two-dimensions the resulting density shows multiple separated modes.

\section{Importance weighted autoencoder test images}

\begin{figure}[!ht]
\centering
\includegraphics[width=0.9\textwidth]{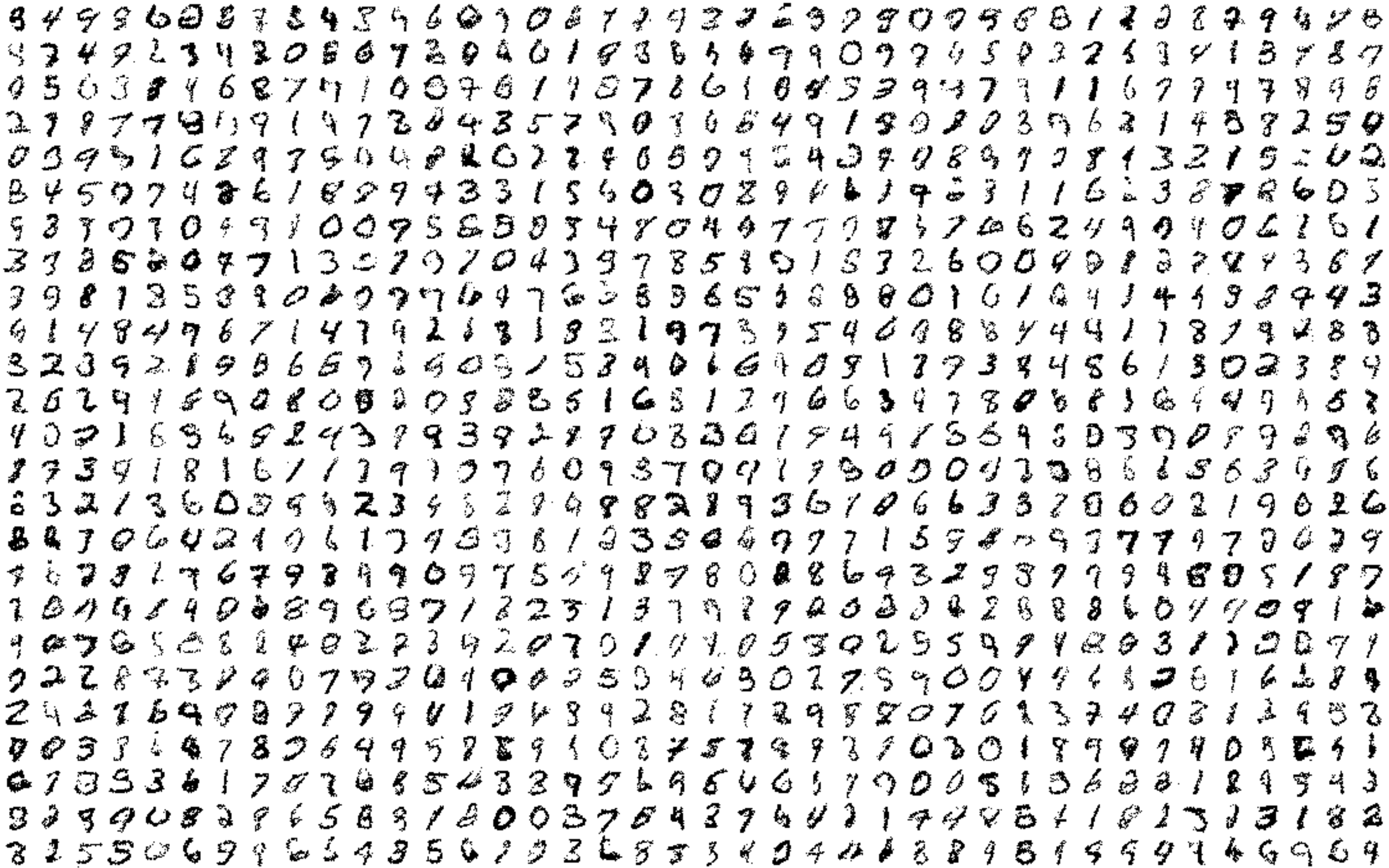}
\caption{MNIST test images.}
\label{fig:mnist-samples}
\end{figure}

\begin{figure}[!ht]
\centering
\includegraphics[width=0.9\textwidth]{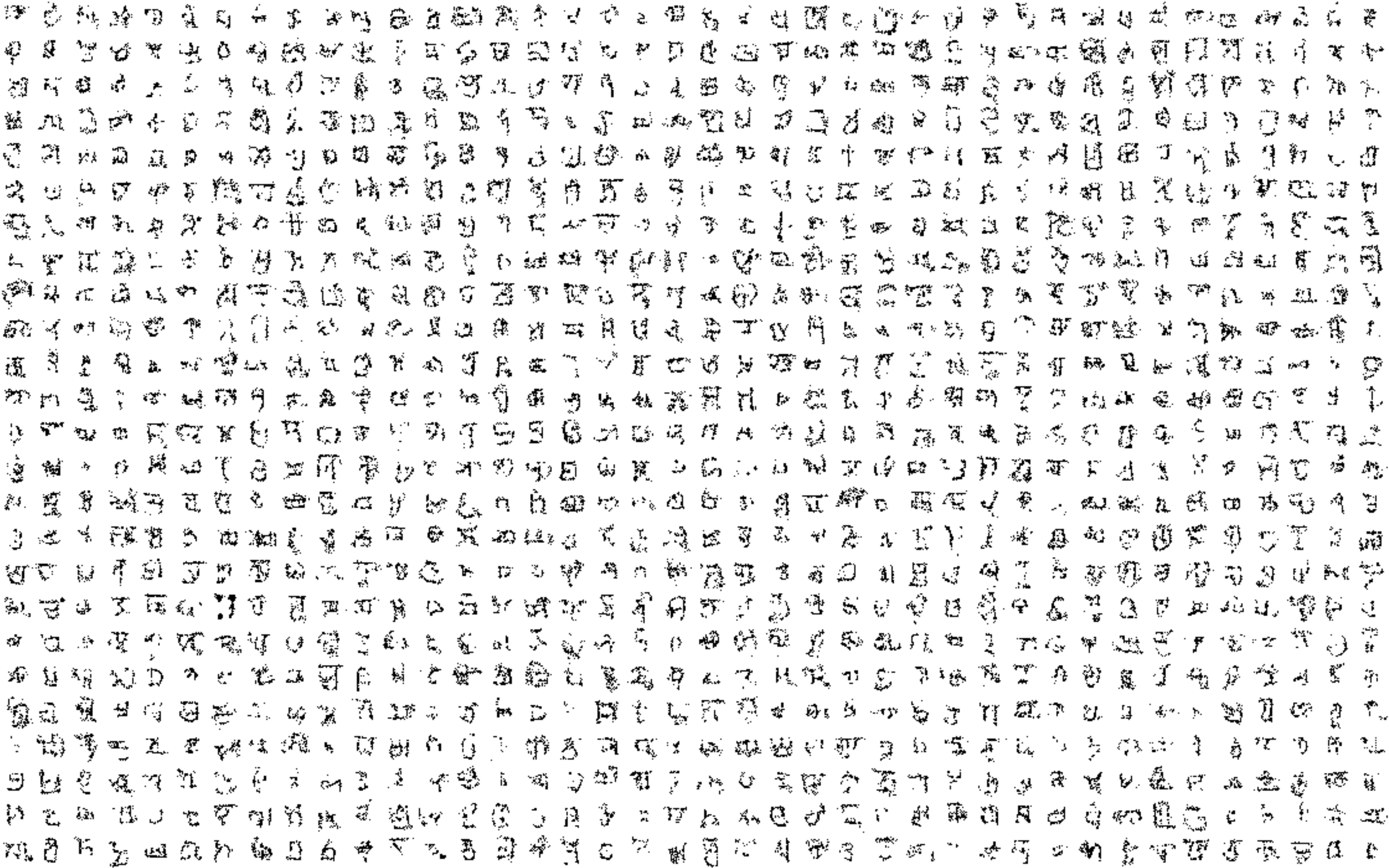}
\caption{Omniglot test images.}
\label{fig:omni-samples}
\end{figure}

\end{document}